\documentclass[11pt,aps,prd,tightenlines,floatfix]{article}
\usepackage[blocks]{authblk}
\usepackage{amsmath,amssymb,graphicx,url}
\usepackage{color}
\newcommand{\half}{ {\scriptstyle \frac{1}{2} } }
\newcommand{\Half}{ {\frac{1}{2} } }

\newcommand\be{\begin{equation}}
\newcommand\ee{\end{equation}}
\newcommand\bea{\begin{eqnarray}}
\newcommand\eea{\end{eqnarray}}
\newcommand{\bdm}{\begin{displaymath}}
\newcommand{\edm}{\end{displaymath}}
\newcommand\nn{ \nonumber\\}

\newcommand{\<}{\langle}
\renewcommand{\>}{\rangle}

\newcommand{\Pm}{P_-}
\newcommand{\Pp}{P_+}

\makeatletter
\def\underbracket{\@ifnextchar [ {\@underbracket} {\@underbracket [\@bracketheight]}}
\def\@underbracket[#1]{\@ifnextchar [ {\@under@bracket[#1]} {\@under@bracket[#1][0.4em]}}
\def\@under@bracket[#1][#2]#3{%\message {Underbracket: #1,#2,#3}
           \mathop {\vtop {\m@th \ialign {##\crcr $\hfil \displaystyle {#3}\hfil $%
                              \crcr \noalign {\kern 3\p@ \nointerlineskip }\upbracketfill {#1}{#2}
                              \crcr \noalign {\kern 3\p@ }}}}\limits}
\def\upbracketfill#1#2{$\m@th \setbox \z@ \hbox {$\braceld$}
                  \edef\@bracketheight{\the\ht\z@}\bracketend{#1}{#2}
                  \leaders \vrule \@height #1 \@depth \z@ \hfill 
                  \leaders \vrule \@height #1 \@depth \z@ \hfill \bracketend{#1}{#2}$}
\def\bracketend#1#2{\vrule height #2 width #1\relax}

\def\overbracket{\@ifnextchar [ {\@overbracket} {\@overbracket[\@bracketheight]}}
\def\@overbracket[#1]{\@ifnextchar [ {\@over@bracket[#1]}{\@over@bracket[#1][0.3em]}}
\def\@over@bracket[#1][#2]#3{%\message {Overbracket: #1,#2,#3}
\mathop {\vbox {\m@th \ialign {##\crcr \noalign {\kern 3\p@\nointerlineskip }\downbracketfill {#1}{#2}
                              \crcr \noalign {\kern 3\p@ }
                              \crcr  $\hfil \displaystyle {#3}\hfil $%
                              \crcr} }}\limits}
\def\downbracketfill#1#2{$\m@th \setbox \z@ \hbox {$\braceld$}
                  \edef\@bracketheight{\the\ht\z@}\downbracketend{#1}{#2}
                  \leaders \vrule \@height #1 \@depth \z@ \hfill
                  \leaders \vrule \@height #1 \@depth \z@ \hfill\downbracketend{#1}{#2}$}
\def\downbracketend#1#2{\vrule depth #2 width #1\relax}
\makeatother

\makeatletter
\@addtoreset{equation}{section}
\makeatother

\begin{document}

\title{The  M\"obius Domain Wall Fermion Algorithm}

\author{Richard~C.~Brower}
\affil{Department of Physics, Boston University,
  590 Commonwealth Avenue, Boston, MA 02215, USA}
\author{Harmut Neff}
\affil{Chamerstrasse 44, 6300 Zug, Switzerland,
email: hartmutneff@aol.com}
\author{Kostas Orginos}
\affil{ Department of Physics,College of William and Mary,Williamsburg,VA 23187-8795}
\affil{Jefferson Laboratory, 12000 Jefferson Avenue, Newport News, VA 23606 }

\maketitle

\begin{abstract} 
  We present a review of the properties of generalized domain wall Fermions, based
  on a (real) M\"obius transformation on the Wilson overlap kernel,
  discussing their algorithmic efficiency, the degree of explicit chiral
  violations measured by the residual mass ($m_{res}$) and the
  Ward-Takahashi identities. The M\"obius class interpolates between  Shamir's domain wall
operator and Bori\c{c}i's domain wall implementation of Neuberger's overlap operator without increasing the number of
  Dirac applications per conjugate gradient iteration. A new scaling parameter
  ($\alpha$) reduces chiral violations at finite fifth dimension ($L_s$) but
  yields exactly the same overlap action in the limit $L_s \rightarrow
  \infty$. Through the
  use of 4d Red/Black preconditioning and optimal tuning for the
  scaling $\alpha(L_s)$, we show that chiral symmetry violations are
  typically reduced by an order of magnitude at fixed $L_s$. At large
$L_s$ we argue that the observed  scaling for  $m_{res} = O(1/L_s)$ for Shamir is replaced
by $m_{res} = O(1/L_s^2)$ for the properly tuned M\"obius algorithm with $\alpha = O(L_s)$.
%  Alternatively  the     M\"obius algorithm  allows
% $L_s$ to be reduced leading  to substantially acceleration the Dirac inverter in lattice evolution and analysis codes at fixed residual %mass.
\end{abstract}

%\pacs{11.15.Ha, % Lattice gauge theory 
%      11.30.Rd, % Chiral symmetries
%     12.38.Aw, % General properties of QCD (dynamics, confinement, etc.)
%      12.38.-t  % Quantum chromodynamics
 %     12.38.Gc  % Lattice QCD calculations
%}

%\newpage
%\setcounter{tocdepth}{3}
%\tableofcontents

%%%%%%%%%%%%%%%%%%%%%%%%%%%%%% INTRODUCTION %%%%%%%%%%%%%%%%%%%%%%%%%%
%\newpage
\section{Introduction}
\label{sec:intro}

Perhaps the most important theoretical  development in lattice field theory at the
end of the last century was the discovery of Fermion actions that
respect exact chiral symmetry at finite lattice spacing.  The
consequence of this is a clean separation between chiral symmetry
breaking (i.e.\ non-zero quark masses) and Lorentz breaking (i.e.\
non-zero lattice spacing). Even at finite lattice spacing, this allows
for a rigorous understanding of topology and greatly simplifies the
numerical extrapolations to obtain renormalized correlation functions
in the continuum limit with the light quarks masses at their physical value.

The key idea in evading the Nielsen-Ninomiya no-go
theorem~\cite{Nielsen:1980rz,Nielsen:1981xu}, which forbade the
construction of lattice Fermion action with chiral symmetry under
rather general conditions, was introduced by
Kaplan~\cite{Kaplan:1992bt}. In his construction four dimensional
chiral zero modes appeared as bound states on a mass defect or 3-brane
in a five dimensional theory. Much like the work of Callan and
Harvey~\cite{Callan:1984sa} in the continuum, anomalous currents in
the 4 dimensional theory are understood as the flow on or off the mass
defect of conserved 5 dimensional currents. This work led to two
concrete realizations of lattice Fermions with chiral symmetry: the
domain wall Fermions~\cite{Shamir:1993zy,Furman:1995ky} and the
overlap
Fermions~\cite{Narayanan:1992wx,Narayanan:1993ss,Narayanan:1994sk,Neuberger:1997bg,Neuberger:1997fp}.
The domain wall formulation retains the five dimensional nature of the
original idea, while the overlap is a direct construction of a four
dimensional effective action. It was subsequently realized, that the
4d effective Fermion operator, $D_{ov}(m)$, of the domain wall action
also satisfies the so called Ginsparg-Wilson (GW)
relation~~\cite{Ginsparg:1981bj},
\be
\gamma_5  D^{-1}_{ov}(0) +   D^{-1}_{ov}(0) \gamma_5  = 2 R \gamma_5 \; ,
\ee
where $D_{ov}(0)$ is the massless Dirac operator and $R = O(a)$ is a
local operator. Taking $R = 1$ in lattice units, a solution to the GW
relation is $D_{ov}(0) = \Half + \Half \gamma_5 \epsilon[ H_5]$ where
$H_5$ is the Dirac Hamiltonian for the 5-th dimension. Clearly the on-shell
 zero mass Fermion action is chiral: $\{\gamma_5,
D^{-1}_{ov}(0)\} \sim 0$.  The Ginsparg-Wilson relation was
discovered in the context of an exact renormalization group improved
Fermion actions in the free field limit~\cite{Ginsparg:1981bj}.  More
than a decade later, it was realized~\cite{Hasenfratz:1997ft}, that
the classical (or so called perfect action~\cite{Bietenholz:1995cy,Bietenholz:1996pf}) approximation to the
renormalization group for the Fermion actions in a fixed gauge
background still satisfies the GW relation.  The Ginsparg-Wilson
relation implies a modified lattice form of infinitesimal chiral
rotations and it can be viewed as an alternative explanation of how the
conditions of the Nielsen-Ninomiya theorem are
circumvented~\cite{Luscher:1998pq}.

Ultimately the domain wall and overlap Fermions are equivalent,
requiring numerical algorithms which implement an approximation to the
GW relations. In the domain wall formulation, this approximation is a
consequence of the finite extent of the fifth dimension, $L_s <
\infty$, whereas for the overlap algorithm a finite rational
approximation to the sign function, $\epsilon_{L_s}[H_5]$. In both
cases the resulting lattice actions have small chiral symmetry
violations parameterized by the defect in the GW relation~(see Eq.~\ref{eq:Delta} below), which in principle can be reduced as needed to
levels that do not seriously affect the physics.

The M\"obius generalization for domain wall Fermions was introduced 7
years ago~\cite{Brower:2004xi,Brower:2005qw} but is just now beginning
to make its way into large scale simulations. Not only has this improvement
not been fully exploited, there are new opportunities to leverage this
algorithm with subsequent developments such as Gap
Fermion~\cite{Brower:2009sb} and as a preconditioner as, for example,
in the MADWF algorithm~\cite{Yin:2011sz}  using Pochinsky's MDWF
inverter code~\cite{Pochinsky:2008zz} or multigrid
methods~\cite{Babich:2009pc,Babich:2010qb, Cohen:2012sh}.  Consequently it is timely to give a review of
some of the basic formalism that leads to the M\"obius generalization of
domain wall Fermions. In addition, the M\"obius generalization calls
for a more careful arrangement of the standard analysis of domain wall
Fermion properties.  While very little of
this formalism is entirely original~\cite{Narayanan:1992wx,Narayanan:1993ss,Narayanan:1994sk,Neuberger:1997bg,Neuberger:1997fp,Edwards:1999bm,Kikukawa:1999sy,Borici:1999da,Borici:1999zw,Brower:2004xi,Brower:2005qw,Borici:2005xxx,Allkoci:2005wd}, we seek to
abstract the basic structure of the domain wall operators and its
equivalence to Overlap in a higher level framework so that the
M\"obius class of generalized domain wall operator is only visible at
the level of implementation much in the spirit of object oriented
software practices.  Consequently we will define precisely the mapping
between the 5d domain wall action and the effective 4d Overlap
operator at fixed $L_s$, including the general way to formulate
residual chiral breaking for correlators and the Ward-Takahashi
identities.

 The  M\"obius generalization introduces two new parameters in the domain wall fermion action. One parameter controls the kernel of the resulting Overlap operator allowing for a continuous family of kernels that have the 
 Neuberger~\cite{Neuberger:1997bg,Neuberger:1997fp} and the Shamir kernels~\cite{Shamir:1993zy} as special cases.   
  The other is a scaling parameter $\alpha$ for the kernel that drops out of the 4D action in the $L_s \rightarrow \infty$ limit, providing   a better approximation to the GW relation at finite $L_s$.
   In this sense it is proper to consider the scaling parameter $\alpha$ as offering a better {\bf
  algorithm} for the same operator.  For finite $L_s$ the result is
that the M\"obius generalization offers a very substantial algorithmic
advantage at fixed residual mass, particularly at large $L_s$.  For example in current practice to
get a sufficiently small residual mass for
thermodynamics~\cite{Cheng:2009be} and ${\cal N} =1$ SUSY~\cite{Giedt:2008xm} the
Shamir algorithm with a fifth dimension as large as $L_s = 48$ has been used.  Here a
rescaled M\"obius gives to reasonably high accuracy an equivalent
action by rescaling to $L_s = 12-16$ or a rescaling parameter $\alpha
= 3-4$.  By reducing the extent of the $5^{th}$ dimension, the Dirac
solver should be accelerated by  roughly this factor $\alpha$ in these
applications. Empirical results are presented here to support this
 estimate.  In addition, it is even possible to further reduce
$L_s$ at fixed $m_{res}$, as noted in the conclusion (see
Figs.\ref{Fig:scaling16} and \ref{Fig:moebius_gap}), by combining the
M\"obius algorithm with Gap method~\cite{Brower:2009sb}.  Consequently the
M\"obius algorithm can very substantially accelerate inverters in
analysis and evolution code with speed up depending on the degree of
chiral symmetry sought.

Finally we recognize interest in combining algorithms based on the equivalence of 
overlap and domain wall formalisms,  since each maybe optimal at different stages of a
single Monte Carlo calculation. To accomplish this one needs to establish
a precise equivalence between the domain wall and overlap formulations
at finite lattice spacing and finite separation $L_s$ of the domain
walls. Here we extend to the M\"obius implementation the exact mapping between the domain wall and overlap Fermion correlators , 
\be
 \< {\cal O}[ q,  \overline  q] \>_{DW}  =  \< {\cal O}[ \psi, \overline  \psi]
 \>_{ov} \;.
%\label{eq:corrId}
\ee
We refer to this  mapping as the {\bf DW/Overlap Correspondence}.
Of course for the Shamir action this mapping has been dealt with
extensively in the literature~\cite{Narayanan:1992wx,Narayanan:1993ss,Narayanan:1994sk,Neuberger:1997bg,Neuberger:1997fp,Edwards:1999bm,Kikukawa:1999sy,Borici:1999da,Borici:1999zw}.
Also we take the opportunity to develop a consistent notation for this
correspondence, extending it to vector and axial currents and their
Ward-Takahashi identities. By studying this for finite $L_s$, we  identify in both domain wall and overlap formulation the 
consequence of violating chiral invariance due to approximations (at
finite $L_s$) that result in an imperfect realization of the
Ginsparg-Wilson identity. The correspondence is based on the descent
relations of Callan and Harvey~\cite{Callan:1984sa} that motivated
Kaplan's original idea. In the 5d space there are only vector currents
since 4d parity becomes a reflection of the 5-th direction. The
symmetric and anti-symmetric projections in the 5-th direction for the
5d vector current become the effective 4d vector and axial currents
respectively of the effective overlap action.

In this paper we separate the general analysis from the
specific details that depend on the choice of the (domain wall, overlap)  action and its
approximation   algorithms (using for example rational polynomial approximation,  domain wall, etc.). In Sec.~\ref{sec:NEF} we
begin by a general statement of the choice of actions and the impact on
an approximation to the Ginsparg-Wilson condition for chiral Fermions.
Following this is the specific domain wall action for the M\"obius Fermion
generalization. Sec.~\ref{sec:RedBlack} explains
the necessity of a 4d Red Black preconditioning for the M\"obius Fermion which is as effective
as the 5d Red Black scheme used for the Shamir variant.  In Sec.~\ref{sec:NumTests} we present the results from  detailed numerical tests demonstrating the effectiveness of our formulation in reducing the cost of implementing domain wall Fermions. Sec.~\ref{sec:equiv} presents the general mapping between
a domain wall action and the overlap action that proves their equivalence at
finite lattice spacing and given rational approximation. The map is applied
to construction of the 4d vector and axial vector currents  form the 5d vector domain wall current.
In Sec.~\ref{sec:WTid} we apply the equivalence formalism to establish the Ward-Takahashi identities
for vector and axial current operators as well as derive the axial and vector currents for M\"obius Fermions. 
In addition, in this section   we construct from the axial ward identity the residual mass as a measure of the violation of chiral symmetry due to 
finite $L_s$ approximation to the overlap operator and related it to the violation of the Ginsparg-Wilson relation.
% The remaining sections
%give some details on applying numerical implementations. This is followed by
%detailed numerical comparisons M\"obius Fermion on quenched $\beta = 6.0$ lattices.

%\newpage
%%%%%%%%%%%%%%%%%%%%%%%%%%%%  Section %%%%%%%%%%%%%%%%%%%%%%%%%%%%%%%%

\section{M\"obius Domain Wall Fermions }
\label{sec:NEF}

%Bori\c{c}i~\cite{Borici:1999da,Borici:1999zw} and others.

Domain wall and overlap Fermions may be viewed as two alternative algorithms for generating chiral Fermions that satisfy the Ginsparg-Wilson relation. This  mapping between domain wall and overlap Fermions is  even useful when chirality is approximated by domain wall Fermions with finite separation between the walls in the $5^{th}$ axis. Throughout this review, we shall emphasize this equivalence by identifying the effective 4d overlap operator,
%y
\be \label{eq:GWapprox}
D^{(L_s)}_{ov}(m) = \frac{1+m}{2} + \frac{1-m}{2}  \gamma_5 \epsilon_{L_s}[\gamma_5 \; D^{kernel}(M_5)] \; ,
\ee
 resulting from the corresponding domain wall implementation at finite extent $L_s$ for the $5^{th}$ dimension. Strictly speaking we should always designate this approximation to the overlap operator by $D^{(L_s)}_{ov}(m)$  but to simplify the notation we leave
this dependence on $L_s$ implicit throughout. The class of
suitable overlap operators is quite large, dictated by the 
choice of the 4d  kernel, $D^{kernel}(M_5)$, and a
particular algorithm to approximate the sign function:
$\epsilon_{L}[x] \simeq \epsilon[x] = x/|x|$.  For example the
standard domain wall implementation gives the polar
approximation~\cite{Vranas:1997da,Kikukawa:1999sy,Edwards:2000qv} to
the sign function  (see Fig.~\ref{Fig:fig6}) ,
\begin{equation}
  \epsilon_{L_s} [H_5] = \frac{(1+H_5)^{L_s}  - (1-H_5)^{L_s}}
                             {(1+H_5)^{L_s}  + (1-H_5)^{L_s}}  \,,
% =   \tanh[- (L/2)\log T]
\label{eq:ApproxEPS}
\end{equation}
where the $5^{th}$ time ``Hamiltonian'' is $H_5 = \gamma_5D^{kernel}(M_5)$
with transfer matrix $T = (1 - H_5)/(1+ H_5)$. 
There are many other ways to generate polynomial or rational
polynomial approximations~\cite{Kennedy:2006ax} to the sign function, some of which have a
natural representation as a local action in the $5^{th}$ time. The
$5^{th}$ time interpretation of Neuberger and Kaplan provide a very
intuitive method to impose chiral symmetry, as we will see when
discussing the axial current. Only in the limit $L_s \rightarrow
\infty$, does the effective domain wall operator reproduce exactly
Narayanan-Neuberger's overlap
Fermion~\cite{Narayanan:1992wx,Narayanan:1993ss,Narayanan:1994sk}.

The exact lattice chiral symmetry (at $m = 0$ and $L_s = \infty$)  is guaranteed by
the G-W relation,
\be
\gamma_5D_{ov}(0) + D_{ov}(0)\gamma_5 = 2 D_{ov}(0) \gamma_5 D_{ov}(0)  \, .
\ee
We may recast G-W relations as, $\gamma_5D_{ov}(0) + D_{ov}(0)
\widehat \gamma_5 =0 $ , by introducing a new $\gamma_5$ operator,
\be
\widehat \gamma_5 =\gamma_5 (1 - 2 D_{ov}(0)) = - \epsilon[H_5] \,,
\label{eq:gamma5hat}
\ee
which   allows one to realize the infinitesimal
chiral transformation  on quark fields as $ \delta \overline \psi = \overline \psi
\gamma_5 , \delta \psi = \widehat \gamma_5 \psi $.  Although the $\widehat
\gamma_5$ depends non-locally on the gauge fields, since the
G-W is equivalent to $\widehat \gamma^2_5 = 1$, it does allow an
unambiguous definition of chiral projection operators:
$\psi_{R/L} = \frac{1}{2}( 1 \pm \widehat \gamma_5) \psi$. 

For finite $L_s$ the  violation of chiral symmetry is given by the error, $\Delta_{L_s}[H_5]$,  in the
Ginsparg-Wilson relation:
\be
 2 \gamma_5 \Delta_{L_s}[H_5]   \equiv \gamma_5 D_{ov}(0) + D_{ov}(0) \gamma_5  - 2 D_{ov}(0)\gamma_5 D_{ov}(0) = \frac{1}{2} \gamma_5 ( 1 - \epsilon^2_L[H_5]) 
\; .
\label{eq:Delta}
\ee
From Eq.~\ref{eq:gamma5hat} this is equivalent to $\Delta_{L_s}[H_5] = (1 + \hat \gamma_5)(1 - \hat \gamma_5)/4$, 
which is a natural way to  measure the failure to implement exactly the projectors: $\hat P_\pm = (1 \pm \hat \gamma_5)/2$. 
This {\bf GW chiral violation operator}, $\Delta_{L_s}[H_5]$, is the unique measure of  chiral symmetry violations  applicable to any  (not necessarily a domain wall) scheme to implement chirality. 
 
 To emphasize this point further, consider how global chiral symmetry of the
Fermion action,
\be 
S^F_{ov} = \sum_{xy} \overline \psi_x D_{ov}(m)_{xy}\psi_y =  \overline \psi_x D_{ov}(0)_{xy}
\psi_y + m \overline \psi_x (1 - D_{ov}(0) ) \psi_x  \; ,
\label{eq:FermionAction}
\ee
is  modified both by the implicit violation due to 
an approximate overlap action and by the explicit breaking
by the bare quark mass term, $m$. The mass term~\footnote{We note that
  the physical range for the mass is $0 \le m < 1$. Indeed it is
  tempting therefore to rescale the action by $(1-m)^{-1} S^F_{ov} =
  \overline \psi_x D_{ov}(0)_{xy} \; \psi_y + m_q \overline \psi_x
  \psi_x$ defining additive bare quark mass $m_q = m/(1-m)$ since the
  limit $m_q \rightarrow \infty$ (or $m = 1$) corresponds to the
  decoupling of the Domain wall quarks by cancellation with
  Pauli-Villars operator $D_{PV} = D_{ov}(m =1)$. However the
  conventional choice has the advantage of introducing the mass
  operator as the correct scalar partner $S(x)$ to the pseudo-scalar
  density $P(x)$.} is given in terms of the scalar density,
\be
 S(x) = \overline \psi_x (1 - D_{ov}(x)) \psi_x  = \overline \psi_x \frac{1 + \gamma_5 \widehat \gamma_5}{2} \psi_x \; ,
\ee
paired with the pseudo-scalar, 
\be
P(x) = \overline \psi_x \gamma_5 (1 - D_{ov}(x)) \psi_x = \overline \psi_x \frac{\gamma_5  + \widehat \gamma_5}{2} \psi_x \; ,
\ee
under the chiral transformation: $\delta S(x) = 2 P(x), \delta P(x) =
2 S(x)$. At finite $L_s$ following the analysis in
Ref.~\cite{Kikukawa:1999sy}, we continue to define the chiral
transformation using $\widehat \gamma_5 = \gamma_5 (1 - D_{ov}(0))$.
The chiral transformation of the action in Eq.~\ref{eq:FermionAction}
now yields two contributions,
\be
\delta S^F_{ov} =  \overline \psi[\gamma_5 D_{ov}(m)
+ D_{ov}(m) \widehat \gamma_5 ] \psi= m \overline \psi(\gamma_5  +  \widehat
\gamma_5)\psi +  2 \overline \psi \gamma_5 \Delta_L \psi \; ,
\label{eq:global}
\ee
the first one due to the quark mass and the second one due to the 
 finite $L_s$ approximation, expressed as same GW violation operator,
$\Delta^{xy}_L $, defined above in Eq.~\ref{eq:Delta}. Like the
overlap operator itself, this chiral violating operator,
$\Delta^{xy}_L$, should also fall off exponentially in units of the
lattice spacing so that it may be approximated by an effective
Lagrangian,
\be
 \overline \psi \Delta_L \psi \simeq m_{res} \overline \psi \psi + c_1 \overline \psi \gamma_\mu (\partial_\mu  - i A_\mu) \psi\, 
+\, i\,c_2 \overline \psi \sigma_{\mu\nu} F^{\mu \nu} \psi + c_4 \overline \psi (\partial_\mu  - i A_\mu)^2 \psi + \cdots \,,
\ee
expanding in the lattice spacing.  The terms in increasing dimension  define a   residual mass, $m_{res}$, the 4d wave function renormalization, the  5d operators or the clover term and so on respectively.  In Sec.~\ref{sec:currents}, we will extend this
analysis to examine the local breaking of chiral symmetry in the
Ward-Takahashi identities for the axial current. There we  show  that 
this same operator, $\overline \psi \gamma_5 \Delta_L \psi$,  is
the correction to the divergence for axial current.
%as  noted in the introduction in Eq.~\ref {eq:axialWT}
Its vacuum to pion matrix element is the conventional definition of the residual mass~\cite{Blum:2000kn,AliKhan:2000iv,Aoki:2002vt} .

%  by a  matrix elements, $\<0| \overline \psi \gamma_5 \Delta_L \psi |\pi\>$, 
%is identical to shift in current algebra mass in the Ward-Takahashi identity.

In spite of the exact  mapping between overlap and domain wall Fermions,
unfortunately the most common implementations
 use different kernels.
For  the overlap it is natural to use the Wilson (or Neuberger/Bori\c{c}i) kernel,
\be 
D^{Borici}(M_5) = a_5 D^{Wilson}(M_5) \; ,
\label{eq:Borici}
\ee
while for  the Shamir domain wall implementation~\cite{Shamir:1993zy},  the kernel is
\begin{equation}  
 D^{Shamir}(M_5) = \frac{a_5 D^{Wilson}(M_5)}{2 + a_5 D^{Wilson}(M_5)}  \; ,
\label{eq:Shamir}
\end{equation}
 where we include in the definition  the lattice spacing $a_5$ in the fifth direction, although it is generally  set to  $a_5 =1$  in units of
the space-time lattice.  Both  the Shamir and Neuberger/Bori\c{c}i  kernels are constructed from the Wilson lattice Dirac operator,
\be
D^{Wilson}_{xy}[U_\mu(x),M_5] =  (4+M_5) \delta_{x,y} - 
 \frac{1}{2} \Bigl[  (1 - \gamma_\mu) U_\mu(x) \delta_{x+\mu,y} 
+  (1 + \gamma_\mu) U_\mu^\dagger(y) \delta_{x,y+\mu} \Bigr] \; , 
\label{eq:D^{Wilson}} 
\ee
with a negative mass term $M_5 = O(-1)$.  Unfortunately the Shamir kernel (\ref{eq:Shamir}) is computationally expensive to use in overlap codes.  On the other hand, as first realized by Bori\c{c}i~\cite{Borici:1999da,Borici:1999zw}, the simpler Wilson overlap kernel (\ref{eq:Borici}) can easily be realized as a domain wall action by adding a specific next to nearest neighbor term in the $5^{th}$ direction. 

\begin{figure}[t]
\begin{center}
\includegraphics[angle = 0,width=0.6\textwidth]{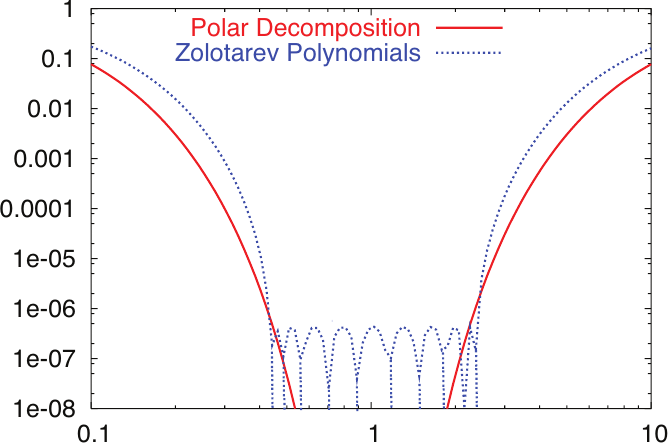}
\caption{  The error, $1 -
  \epsilon_{L_s}(\lambda)$ for the polar decomposition at $L_s = 16$ is
  plotted against an eigenvalue of the kernel $H_5$ for $\lambda \in [0.1,10]$. In the polar approximation the
  error is positive semi-definite for $\lambda \ge 0$, anti-symmetric for $\lambda \rightarrow -\lambda$
  and symmetric under $\lambda  \rightarrow 1/\lambda$.  This is compared with $|1 -
  \epsilon_{L_s}(\lambda)|$ for the Zolotarev polynomial at $L_s = 10$ whose
  error fluctuates in sign.}
\label{Fig:fig7}
\end{center}
\end{figure}

Here we found that an equally computationally efficient domain wall action allows one to introduce
the generalized kernel,
\be
D^{Moebius}(M_5) =  \frac{(b_5 + c_5)D^{Wilson}(M_5)}{2 +  (b_5 - c_5) D^{Wilson}(M_5)} \; ,
\label{eq:Moebius}
\ee
referred to as M\"obius  Fermions because the  3 parameters  $M_5, b_5, c_5$ are equivalent to a real M\"obius transformation, 
$D^{Wilson}(-1) \rightarrow [a + b D^{Wilson}(-1)]/[c + d
D^{Wilson}(-1)]$,  of the Wilson operator -- the most general conformal map preserving the real
axis.  It encompasses both the Shamir and Bori\c{c}i form as special cases.  Obviously the standard polar decomposition for the Shamir kernel
is recovered with $b_5=a_5$ and $c_5=0$ while  the Bori\c{c}i truncated overlap action is implemented
 for $b_5=c_5=a_5$.  In addition to the Shamir parameter, $a_5 =
b_5\! - \! c_5$, there is a new scale factor, $\alpha = (b_5\! + \!
c_5)/a_5 $, which turns out to have a major impact on  reducing chiral symmetry violation at finite $L_s$.
Just rescaling the Shamir kernel we have  
\be
D^{Moebius}(M_5) =  \alpha \frac{a_5D^{Wilson}(M_5)}{2 +  a_5 D^{Wilson}(M_5)} \;  \equiv \; \alpha D^{Shamir}(M_5) \; .
\label{eq:Moebius2}
\ee We will also extend our analysis to s-dependent coefficients, $
a_5(s) = b_5(s) - c_5(s)$, $\alpha(s) a_5(s) = b_5(s) + c_5(s)$, so
that this presentation includes the Zolotarev approximation (for
example) as described in Ref~\cite{Chiu:2002ir,Chiu:2002kj}.  See
Fig.~\ref{Fig:fig7} for an illustration of the error in the sign
function for the polar versus the Zolotarev approximation.

\begin{figure}[ht]
\begin{center}
\includegraphics[width=0.6\textwidth]{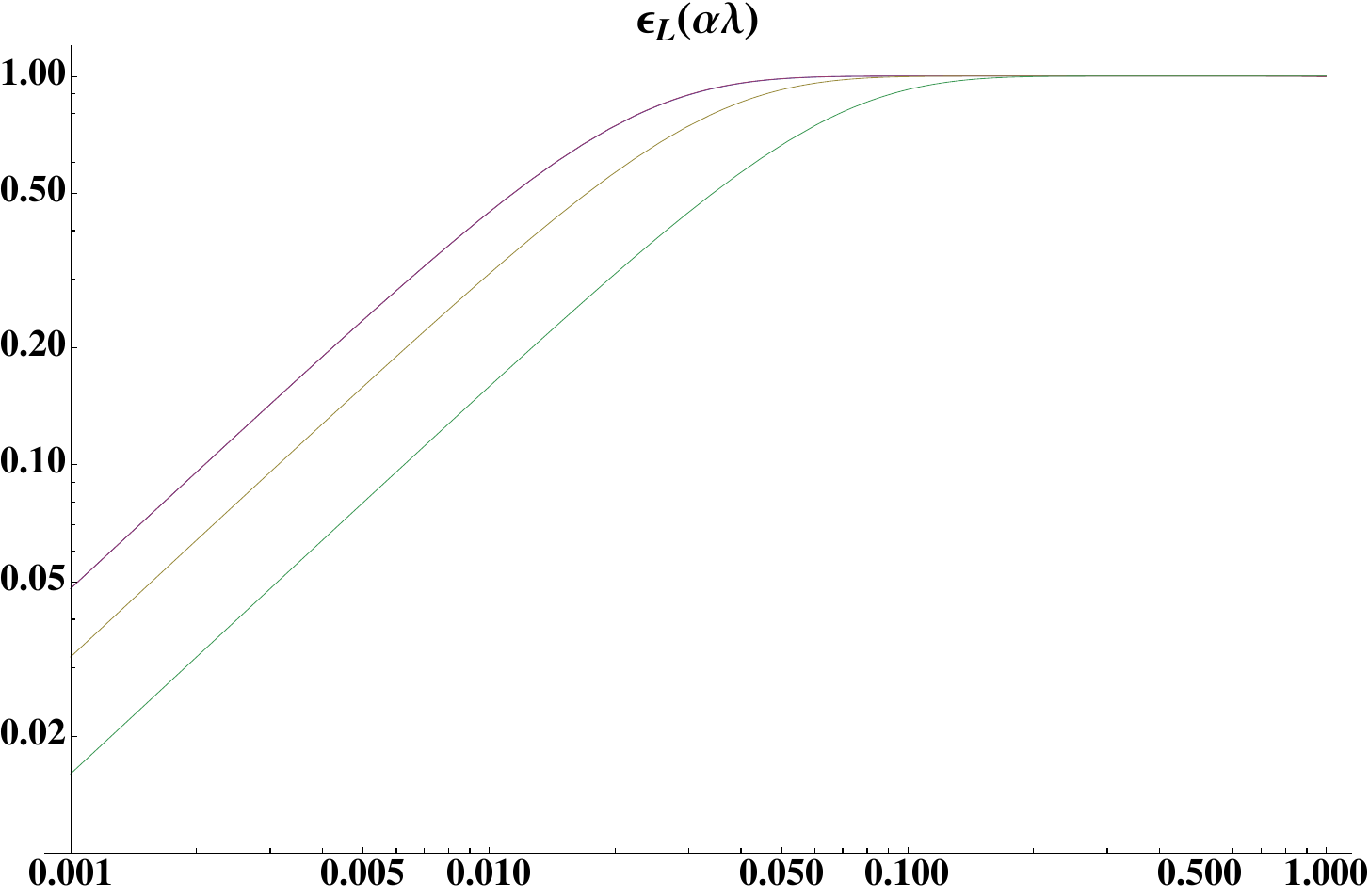}
\caption{  The top curve
represents the sign function $\epsilon_{L_s}(\alpha \lambda)$ plotted against $\lambda$  for 
Shamir  ($\alpha = 1$) at $L_s = 48$  compared
with  scaled M\"obius ($\alpha  = 4$) at   $ L_s = 12$, which
are  indistinguishable on a loglog plot. This contrast with  the middle and lower curves for Shamir at $L_s=32$ and $L_s = 16$ respectively which do degrade the chiral  approximation substantial. }
\label{Fig:scaling}
\end{center}
\end{figure}

The scaling parameter plays a particularly important role in  M\"obius Fermions.  Since
the sign function is scale invariant, if we simply rescale the Shamir action, $\epsilon_{Ls} [H_5]= \epsilon_{L_s/\alpha} [\alpha H_5]$,  the M\"obius action is identical to the Shamir
action in the $L_s \rightarrow \infty$ limit.  Thus it is proper to  regard this as a new algorithm with improved
chirality at finite $L_s$.   A heuristic explanation  for this is easily understood.  Consider how the polar approximation
in Eq.~\ref{eq:ApproxEPS}, 
\be
\epsilon_{L_s}(H_5) = \sum_\lambda | \lambda \> \epsilon_{L_s}(\lambda) \< \lambda |\,,
\ee
in the spectral representation approaches the sign function $\epsilon(H_5)$. 
The  polar approximation, illustrated in Fig.~\ref{Fig:fig7},  is exact at  $\lambda = \pm 1$ and has an error $\Delta_{L_s}(\lambda) = (1 - \epsilon^2_{L_s}(\lambda))/4$ that satisfies the inversion symmetry $\Delta_{L_s}(\lambda) = \Delta_{L_s}(1/\lambda)$.  It converges exponentially for increasing $L_s$ in a window $O(1/L_s) < |\lambda| < O(L_s)$.  However since the spectrum of $H_5$ is bounded $|\lambda| \le \lambda_{max}$  the upper end of the window is wasted whenever $L_s > O(\lambda_{max})$.  Consequently by
 using the M\"obius scaling parameter  $\alpha$, the log spectrum can be shifted
and for $\alpha \lambda_{max} < O(L_s)$ we take advantage of the entire window in the polar approximation, greatly improving the approximation for the small eigenvalues  with essentially no damage to the approximation at the  high end of the eigenvalue spectrum. For example with the Shamir kernel with $a_5 =1$, $\lambda_{max} \simeq (8-M_5)/(10- M_5) \simeq 3/4 $ so we can rescaling the Shamir kernel by  $\alpha = O(L_s)$ allowing the window to cover better the 
low eigenvalues that are mostly responsible for the observed explicit chiral symmetry breaking.
 
In addition note that for low eigenvalues and typical values of $\alpha$ (such as those used in our numerical tests), the sign function obeys the scaling rule, $\epsilon_{L_s}(\lambda) \simeq \epsilon_{L_s/\alpha}(\alpha \lambda)$.
 Indeed  as illustrated in Fig.~\ref{Fig:scaling}, this scaling rule holds quite well 
 for a substantial part of the low end of the spectral range of the  Hermitian kernel $H_5$ so the Domain Wall operator is 
essentially unchanged under this  ``approximate  equivalence relation'':
\be
\mbox{\bf Scaling Rule:} \quad  \mbox{Shamir at} \; \; L_s \simeq  \mbox{M\"obius at} \; \; L_s/\alpha \; .
\label{eq:scaling}
\ee
Since each eigenvector of the operator  and therefore the operator itself is   preserved,   the M\"obius
rescaling can be understood as an improved algorithm for  essentially the same lattice Dirac action.  For codes requiring an aggressive chiral approximation, the
advantage of the M\"obius algorithm will be dramatic.   For example, 
 recently  the Shamir domain simulation  have required $L_s = 48$  for ${\cal N} =1$ SUSY~\cite{Giedt:2008xm} and $L_s = 32$ for thermodynamics~\cite{Cheng:2009be}  or
even larger $L_s$  to get sufficiently small $m_{res}$.     However 
 as demonstrated in Fig.~\ref{Fig:scaling} the rescaling from  Shamir at $L_s =  48$ to 
M\"obius  at $L_s = 12$ and  $\alpha = 4$  has no visible effect on a loglog
plot. The difference is  $|\epsilon_{48}[\lambda]- \epsilon_{12}[ 4 \lambda] | < 10^{-3}$ for all the vectors of $H_5$. Consequently any simulation
using $L_s = 32$ or larger  should benefit very  substantially by  a factor of 2-4 using M\"obius in the range of $L_s =12$ to $16$ with equally small values of $m_{res}$.

\subsection{Domain Wall Action for M\"obius Fermions}

\begin{figure}[ht]
\begin{center}
\setlength{\unitlength}{0.7 mm}
\begin{picture}(220,100)
\linethickness{.35mm}
\put(20,10){\vector(1,0){160}}
\multiput(40,10)(40,0){4}{\dashbox{2}(20,80)}
%\put(0,10){\dashbox{2}(6,80)}
\multiput(20,10)(20,0){1}{\line(0,1){80}}
\multiput(180,10)(20,0){1}{\line(0,1){80}}
\put(20,90){\line(1,0){160}}
%\multiput(0,10)(20,0){6}{\line(0,1){4}}
%\multiput(180,10)(20,0){2}{\line(0,1){4}}
%\put(0,10){\circle{2}}
\put(190,20){\large $q^R = P_+ \Psi_{L_s}$}
\put(-20,20){\large $q^L = P_- \Psi_1$}  
\put(20,0){\large $ 1$}  
\put(40,0){\large $ 2$}  
\put(60,0){\large $ 3$}  
\put(100,0){\Large $s \; \rightarrow $}  
\put(155,0){\large $L_s - 1$}  
\put(180,0){\large$ L_s$}  
\end{picture}
\caption{Domain wall convention  with the physics Right/Left chiral mode
 at $s = 1$ and $s = L_s$ respectively for an approximation
of the fifth dimension by width $a_5 L_s$. The Pauli-Villars operator has
anti-periodic boundary condition and the Dirac operator Dirichlet boundary
condition at zero quark mass. }
\label{Fig:domainwall}
\end{center}
\end{figure}
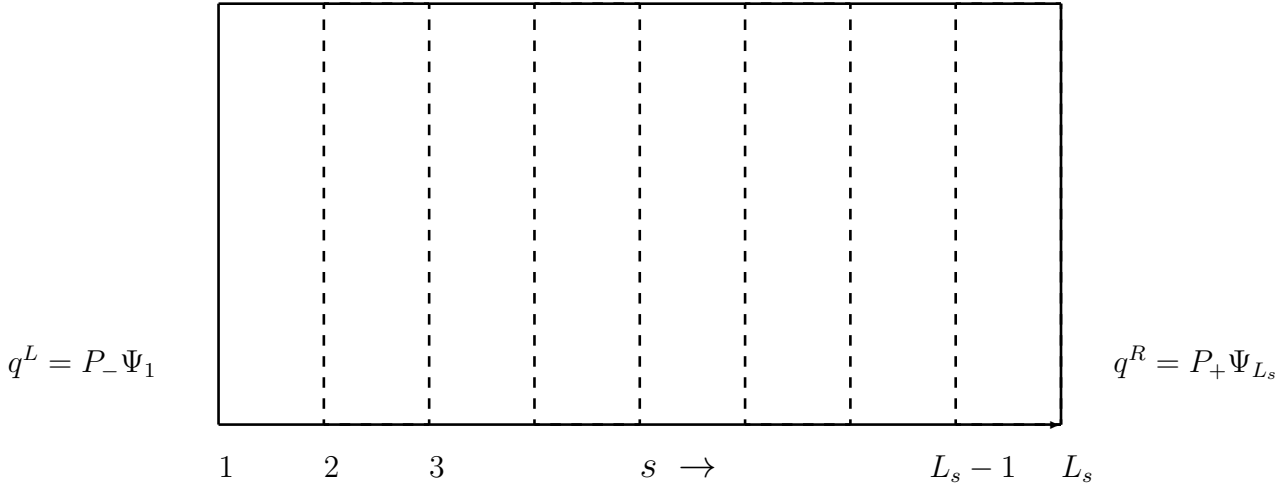

We now proceed to construct the domain wall action that gives
rise to the   M\"obius kernel.  It is useful to introduce two generalized kernels:
\be
D^{(s)}_+ =  b_5(s) D^{Wilson}(M_5) + 1  \quad \mbox{and} \quad D^{(s)}_- = c_5(s) D^{Wilson}(M_5) - 1\,,
\ee
with $s = 1, 2, \cdots L_s$.  
%Fixing 
%$b_5(s) = a_5$,  $D^{(s)}_+$ accounts for the  standard Shamir term generalized to depend on the 
%5-th coordinate and $ D^{(s)}_- $ introduces the non-trivial 5-th dimension hopping term required for the M\"obius extension. 
 We keep an arbitrary  $s$-dependence to 
allow for generalization  of our formulation to other five dimensional approximations to the overlap  such as  the Zolotarev polynomial approximation. Also the s-dependence of  $D^{(s)}_+ $ and $D^{(s)}_- $   allows for generalizations to formulations that require a gauge field dependence of the fifth dimension such as  the domain wall filters suggested
in Ref.~\cite{Bar:2007ew}.   Now our generalized  domain wall Fermion action~\footnote{Through out we adopt the convention
   of upper case for 5d domain wall fields (e.g. $\Psi_{x,s}$) and
   lower case for 4d overlap fields (e.g. $\psi_x$) and 4d domain wall
   fields on the boundary (e.g. $q = [{\cal P}^\dagger \Psi]_1$).} is 
\bea
\overline \Psi D^{DW}(m)\Psi &=& \sum^{L_s}_{s=1} \overline \Psi_s D^{(s)}_+ \Psi_s 
      +    \sum^{L_s}_{s=2}     \overline  \Psi_s   D^{(s)}_- P_+ \Psi_{s-1} 
      +    \sum^{L_s-1}_{s=1}  \overline \Psi_s   D^{(s)}_- P_-  \Psi_{s+1}  \nonumber \\
      &- & \;  m \;  \overline \Psi_1 D^{(1)}_-  P_+\Psi_{L_s} - \;  m \;  \overline \Psi_{L_s} D^{(L_s)}_- P_-\Psi_{1} \,,
\label{eq:DWaction}
\eea
where we follow the conventions of Edwards and Heller~\cite{Edwards:1999bm},
placing the left chiral modes, $q^L = P_- \Psi_1$ on the $s=1$ wall and the
right chiral modes, $q^R = P_+ \Psi_{L_s}$ on the $s=L_s$ wall as depicted in
Fig~\ref{Fig:domainwall}. $P_\pm = \half (1 \pm \gamma_5)$ are the
chiral projectors. Written as an $L_s \times L_s$ tridiagonal matrix the M\"obius operator is 
\be
D^{DW}(m) =
\begin{bmatrix}
D^{(1)}_+ & \quad D^{(1)}_- P_- & 0 &\cdots &  -mD^{(1)}_- P_+  \\
\quad D^{(2)}_- P_+ & D^{(2)}_+ & \quad D^{(2)}_- P_- &\cdots & 0  \\
0 & \quad D^{(3)}_- P_+ & D^{(3)}_+ & \cdots&   0  \\
\cdots & \cdots & \cdots & \cdots &  \cdots  \\
-mD^{(L_s)}_- P_- & 0 & 0  &  \cdots & D^{(L_s)}_+  \\
\label{eq:MoebiusOp}
\end{bmatrix}\,.
\ee
For the polar decomposition, it is sufficient to choose constant
coefficients $b_5(s) = b_5$, $c_5(s) = c_5$ for all $s$ with the kernel taking the
form,
\be
D^{Moebius}(M_5) = \frac{(b_5 + c_5)D^{Wilson}(M_5)}{2 +  (b_5 - c_5) D^{Wilson}(M_5)} =  \frac{D_+ + D_-}{D_+ - D_-} \; .
\ee
It is worth noting that there maybe a computational advantage to placing the chiral projectors, $P_\pm$, to the left of $D^{(s)}_-$.  Converting to this form is a trivial similarity transformation, $D^{DW}(m) \rightarrow \widehat D^{DW}(m) = D^{-1}_- \; D^{DW}(m) D_- $ where we introduce the matrix,
$$ D_- = Diag\left[\begin{array}{ccccc}
D^{(1)}_- & D^{(2)}_- &  D^{(3)}_- &\cdots & D^{(L_s)}_-\\
\end{array} \right] \; ,$$
diagonal in the $5^{th}$ axis.  Assuming reflection symmetry
$ D^{(s)}_- = {\cal R}_{ss'} D^{(s')}_- = D^{(L_s + 1 -s)}$ the generalization of  $\gamma_5$-Hermiticity is 
\be 
\Gamma_5  D^{DW}(m) =  D^{\dag DW }(m) \Gamma_5 \,.
\ee
Here  $\Gamma_5 = \gamma_5 {\cal R} D^{-1}_-$  is defined to include inverting the 5-th axis,
\be
{\cal R}_{ss'} = \delta_{L_s + 1 - s, s'}\,,
\label{eq:Rsym}
\ee
 and rescaling by $ D^{-1}_-$.  The fact that $H^{DW} = \gamma_5 {\cal
  R} D^{-1}_- D^{DW}(m)$ is a Hermitian operator
 guarantees  that the effective 4d overlap operator is ``$\gamma_5$ Hermitian''~\footnote{In terms of
  the transfer operator on the $(s+1,s)$ link introduced later this
  condition is $T_s = T_{L_s +1 -s}$. In the case of the Zolotarev approximation,  one uses the condition that
  s-dependent transfer matrix commutator vanishes ($[T_s ,T_{s'}] = 0$) to show  $\gamma_5$ Hermiticity of the 4d effective operator.}.

\subsection{Domain Wall to  Overlap Reduction}

The equivalence between  domain wall and overlap Fermions (which has been established in the literature as discussed in the introduction)
is constructed  by dimensional reduction from 5 to 4 dimensions~\footnote{The details of the formulation discussed here first appeared in~\cite{Brower:2004xi}.}
Since the Fermion action is quadratic,  there
is an explicit map to the 4d overlap operator from the 5d domain wall action.  
To accomplish this is a straight forward application of
{\bf LDU decomposition} for both the Dirac, ($D^{DW}(m){\cal P}$) and
Pauli-Villars ($D^{DW}(1) {\cal P}$) matrices, as demonstrated in
Appendix~\ref{sec:A}.  This procedure  makes use of a  permutation matrix,
\begin{equation}
 {\cal P} =  \left[ 
{\begin{array}{cccc}
\Pm & \Pp & \cdots & 0 \\
0 & \Pm & \Pp \cdots &0 \\
\vdots & \vdots & \ddots & \vdots \\
0 & 0 & \cdots & \Pp\\
\Pp & 0 & \cdots & \Pm
\end{array}}
 \right] \; ,
\end{equation}
This matrix,  $ {\cal P}_{ss'} = \delta_{ss'} \Pm +
\delta_{s',(s+1)\, mod\, L_s} \Pp $, is a unitary permutation,
${\cal P}^{-1} = {\cal P}^{\dagger} ={\cal R} {\cal P} {\cal R}$, which
rotates the positive chiral
mode at $s = L_s$ to the same position as the left chiral mode at $s =
1$, so that the new 4d Dirac field is reassembled at $s=1$:
\be
q_x = P_- \Psi_{x,1} + P_+ \Psi_{x,L_s} = [{\cal P}^\dagger
\Psi]_{x,1} \; .
\ee

The main result of the LDU  decomposition in Appendix~\ref{sec:A}
is that we are led to consider the domain wall Dirac operator
preconditioned by the Pauli-Villars   (corresponding to infinite quark  mass) operator,
\be
K^{DW}(m) = {\cal P}^{\dagger} \frac{1}{D^{DW}(1)}D^{DW}(m) {\cal P}  \; .
\ee
This matrix has a remarkably simple form  given by
\begin{equation}
K^{DW}(m) =  {\cal P}^{\dagger} \frac{1}{D^{DW}(1)}D^{DW}(m) {\cal P} = 
 \left[ \begin{array}{rrrrrrr}
D_{ov}(m) & 0 & 0 & \cdots & \cdots & \cdots& 0\\
-(1-m)  \Delta^R_{2}  & 1 &0 &0&\cdots &\cdots& 0\\ 
-(1-m)\Delta^R_{3} & 0 &1 & 0 & \cdots &\cdots&0\\
-(1-m)\Delta^R_{4} & 0 & 0 & 1  &  & \cdots & 0 \\
\vdots & \vdots &  \ddots & \ddots &\ddots &\ddots &\vdots \\
-(1-m)\Delta^R_{L_s} & 0 & \cdots &\cdots&\cdots&0& 1
\end{array}\right]
\label{eq:DWFmatrix} \; .
\end{equation}
Its inverse, or propagator matrix, has the form
\begin{equation}
A^{DW}(m) =   {\cal P}^{\dagger} \frac{1}{D^{DW}(m)}D^{DW}(1) {\cal P} = 
 \left[ \begin{array}{rrrrrrr}
D^{-1}_{ov}(m) & 0 & 0 & \cdots & \cdots & \cdots& 0\\
(1-m)\Delta^R_{2} D^{-1}_{ov}(m)  & 1 &0 &0&\cdots &\cdots& 0\\ 
(1-m)\Delta^R_{3} D^{-1}_{ov}(m)  & 0 &1 & 0 & \cdots &\cdots&0\\
(1-m)\Delta^R_{4}  D^{-1}_{ov}(m)  & 0 & 0 & 1 & \cdots & \cdots & 0 \\
\vdots & \vdots &  \ddots & \ddots &\ddots &\ddots &\vdots \\
(1-m) \Delta^R_{L_s}  D^{-1}_{ov}(m)  & 0 & \cdots &\cdots&\cdots&0& 1
\end{array}\right]  \; .
\label{eq:inverseDWFmatrix}
\end{equation}
The results for the $K^{DW}$ matrix in Eq.~\ref{eq:DWFmatrix} and its inverse in  Eq.~\ref{eq:inverseDWFmatrix}  play a fundamental role in  understanding the dynamics of the domain wall Fermion construction. They state the remarkable
fact, that once the domain wall Fermion operator, $D^{DW}(m)$, is
``preconditioned'' by the Pauli-Villars operator, $D^{DW}(1)$, the
propagation into the fifth dimension from the chiral domain wall is
``instantaneous''.    All the essential properties are independent of the particular structure of the domain wall operator.
The ``instantaneous'' property is easily demonstrated
by rewriting the propagator as the sum of two terms,
\be
{\cal P}^{\dagger} \frac{1}{D^{DW}(m)}D^{DW}(1) {\cal P} = 1 + {\cal P}^{\dagger} \frac{1}{D^{DW}(m)}(D^{DW}(1) - D^{DW}(m))  {\cal P} \; .
\ee
The first term contributes exclusively to the non-zero value on the diagonal and the second to the first column,
\be
[(D^{DW}(1) - D^{DW}(m)) {\cal P}]_{s',s}  = -(1 -m) {D^{(s')}_-\cal P}_{s',1} \delta_{s,1} \; ,
\ee
because the mass parameter in the domain wall operator is restricted to the link connecting the left and right walls. The zeros in the above equation represent  the cancellation between bulk modes of the Dirac and Pauli-Villars terms.  In the continuum the cancellation occurs~\cite{Kaplan:1999jn} for all off-diagonal terms  but otherwise is very similar.

The  non-zero elements depend only  on the  transfer matrix along the $5^{th}$ dimension,
\begin{equation}
  T_s = \frac{1-H_s}{1+H_s} \quad \mbox{wit
h} \quad H_s = \gamma_5 
\frac{(b_5(s) + c_5(s)) D^{Wilson}(M_5)}{2+ (b_5(s) - c_5(s)) D^{Wilson}(M_5)} \; .
\end{equation}
The effective overlap operator is then given by
\be
D^{ov}(m) =  \frac{1+m}{2} + \frac{1-m}{2} \gamma_5 \frac{{\mathbb T}^{L_s}-1 }
{{\mathbb T}^{L_s}+ 1 }\,,
\ee
where  ${\mathbb T}^{-L_s} \equiv T^{-1}_1 T^{-1}_{2}\cdots T^{-1}_{L_s}$.
The rest of the first column are built out of partial ordered products 
\be
 \Delta^L_s =  \frac{T^{-1}_1 T^{-1}_{2}\cdots T^{-1}_{s}}{  1 + {\mathbb T}^{-L_s}}  \quad, \quad \Delta^R_{s+1} =  \frac{T^{-1}_{s+1} T^{-1}_{s+2}\cdots T^{-1}_{L_s}}{  1 + {\mathbb T}^{-L_s}}  \; .
\ee
We also note that the  GW violation operator, $\Delta_{L_s}$, in Eq.~\ref{eq:Delta}  has a similar expression joining the left and right wall,
\begin{equation}
  \Delta_{L_s} = \Delta^L_{s} \Delta^R_{s+1}  =  \left[\frac{1}{{\mathbb T}^{-L_s/2}+{\mathbb T}^{L_s/2}}\right]^2 =   \frac{1}{4}\Bigg[1-\epsilon_{L_s}^2(H_5)\Bigg] \; .
\label{eq:Delta_Ls}
\end{equation}
Clearly convergence to the exact chiral limit, $\Delta_{L_s\rightarrow \infty } = 0$, depends on the spectrum of the
transfer operator as is evident in the original construction of the overlap operator by Neuberger et al~\cite{Narayanan:1992wx,Narayanan:1993ss,Neuberger:1997bg}.

\subsection{Overlap to bulk Domain Wall reconstruction}

We can essentially reverse the 5d to 4d dimensional reduction at a small
cost using the more easily inverted Pauli-Villars operator. This is
useful in a variety of instances, the most obvious being the
reconstruction of the full domain wall solution, when you have only
saved the solution to the effective overlap propagator.  Another
recent example is the use of a M\"obius operator at small $L_s$ as a
preconditioner for larger $L_s$ standard Shamir operator, as in the
recent M\"obius Accelerated Domain Wall Fermion (MADWF) algorithm~\cite{Yin:2011sz}.

Let's begin by  assuming  we have the solution to the effective overlap  equation,
\begin{equation}
 D_{ov}(m) \psi = b \; ,
\end{equation}
and proceed to find the domain wall propagator assuming that its
source is given by $B_s = b \; \delta_{s,1} $ on the boundary.  The
goal is to reconstruct the 5d bulk propagator. In the first step using
Eq.~\ref{eq:DWFmatrix} and defining, $ \widetilde \psi_s = \psi\;
\delta_{s,1} $, we reconstruct the vector in the interior, 
\begin{equation}
\Omega  = {\cal P}^{\dagger} \frac{1}{D^{DW}(1)}D^{DW}(m) {\cal P}  \widetilde \psi =
  \left[ \begin{array}{r}
   D_{ov} \psi \\
   -(1-m)  \Delta^R_{2} \psi\\
  -(1-m)\Delta^R_{3} \psi \\
  \vdots \\
  -(1-m)\Delta^R_{L_s} \psi
   \end{array}\right]  \; .
\end{equation}
with the cost of having to invert the Pauli-Villars matrix.
Using Eq.~\ref{eq:inverseDWFmatrix} we  see that the  5d solution $\Psi$  is given by 
\begin{equation}
\Psi =  {\cal P}^{\dagger} \frac{1}{D^{DW}(m)}D^{DW}(1) {\cal P}  B=
  \left[ \begin{array}{r}
   \psi \\
   (1-m)  \Delta^R_{2} \psi\\
  (1-m)\Delta^R_{3}\psi  \\
  \vdots \\
  (1-m)\Delta^R_{L_s}\psi
   \end{array}\right]  \; ,
\end{equation}
where  $B_s = b \; \delta_{s,1}$ is a 5d vector with support on the boundary only.
 
Comparing the two results we have
\begin{equation}
\Psi^{dwf}_s = {\rm diag}( D_{ov}^{-1}, -1,\cdots,-1) \Omega = \left\{
 \begin{array}{lr}
 \psi & s=1\\
 -\Omega_s & s\ne 1
 \end{array}
\right.
\,.
\end{equation}
This final step  only requires  flipping the sign, $\Psi_s = - \Omega_s$ for
$s = 2,\cdots , L_s$. The cost of this construction is the cost of solving a
5d linear system for  the Pauli-Villars matrix which   converges 
in $O(50)$   conjugate gradient iterations for typical applications.

Using this formulation we can construct the conserved axial-vector and vector current matrix elements
using only the stored overlap propagators. Furthermore, we can use this formalism to define prolongation and
restriction operations to coarsen the fifth dimension in the spirit of a multigrid solver.   However
it should be realized that a true domain wall multigrid algorithm~\cite{Cohen:2012sh} requires blocking in the 4d to overcome
the critical slowing down in the chiral limit.

\subsection{4d Hybrid Monte Carlo}

Another related consequence of the fundamental identity in Eq.~\ref{eq:inverseDWFmatrix}  is the ability to reformulate Hybrid Monte Carlo evolution for domain wall Fermions restricting the pseudo-fermions to the wall.  For example consider the standard approach to HMC in using domain wall Fermions in dynamical Fermion calculations~\cite{Izubuchi:2002pt,Antonio:2006px}
\begin{equation}
  \det[D_{ov}^\dagger D_{ov}] = \int d\phi_1\cdots d\phi_{L_s}
e^{ \textstyle -\phi^\dagger {\cal P}^{\dagger} D^{\dagger DW}(1)\frac{1}{D^{DW}(m)^\dagger} \frac{1}{D^{DW}(m)} D^{DW}(1){\cal P}\phi}\,.
\end{equation}
for  2 flavor example that introduces pseudo-fermions $\phi_s$ in the bulk. If we look at the  action $S(\phi^\dagger,\phi)$, 
\begin{eqnarray}
S(\phi^\dagger,\phi)&=& \phi^\dagger_1\frac{1}{D_{ov}^\dagger}\frac{1}{D_{ov}}\phi_1   \nonumber \\
&\; +& \sum_{s=2}^{L_s}  \left[\phi_s+(1-m)\Delta_s^RD^{-1} _{ov}\phi_1\right]^\dagger 
\left[\phi_s+(1-m)\Delta_s^RD^{-1} _{ov}\phi_1\right]  \; ,
\end{eqnarray}
we see that only  the first term contributes to the path integral. The remaining  terms
integrate to unity trivially but they do add unwanted noise to the stochastic estimator.
 
 From the reconstruction
procedure described above  the bulk degrees of freedom should  be redundant.
Indeed this is true if we use the  identity 
\be
  (D^\dagger_{ov}(m) D_{ov}(m))^{-1}  
  =  [A^{DW} A^{DW \dag}]_{11} =  [{\cal P}^\dagger \frac{1}{D_{DW}(m)} D_{DW}(1) D^\dagger_{DW}(1) 
 \frac{1}{D^{\dagger_{DW}(m)}}{\cal P}]_{11}
  \ee
that follows from Eq.~\ref{eq:inverseDWFmatrix} to rewrite the
determinant in terms of pseudoFermions restricted to $s = 1$.   
\begin{equation}
  \det[D_{ov}^\dagger D_{ov}] = \int d\phi_1 
e^{ \textstyle -\phi^\dagger_1  {\cal P}^\dagger \frac{1}{D_{DW}(m)} D_{DW}(1) D^\dagger_{DW}(1) 
 \frac{1}{D^{\dagger_{DW}(m)}}{\cal P}\phi_1}\,.
\end{equation}
The benefit of removing the redundant random pseudo-fermions from the bulk
may be significant.  
%The conventional 5d hybrid Monte Carlo
%scales with $V_{5d}^{5/4}$ and this can be reduced  to  $V_{4d}^{5/4} \times L_s$
%using our  4d HMC with the factor of $L_s$ coming from
%inverting $D^{DW}$ in the force calculation. 
Unfortunately,  we should also  note that the force term now requires two inversions instead of one as seen from the expression, 
\bea
 F_\mu &=& \delta_{\omega_\mu} \left[ \phi^\dagger  {\cal P}^\dagger D^{-1}_{DW}(m)
D_{DW}(1) D^\dagger_{DW}(1) 
  D^{\dagger -1}_{DW}(m){\cal P}  \phi\right]  \nonumber \\
&=& \chi^\dagger \delta_{\omega_\mu}\left[ D_{DW}(m) \right]   \psi +  
 \psi^\dagger \delta_{\omega_\mu}\left[ D^\dagger_{DW}(m) \right]    \chi + 
\chi^\dagger \delta_{\omega_\mu}\left[ D_{DW}(1) D^\dagger_{DW}(1) \right] \chi  \; ,
\eea
 in terms of auxiliary $\chi$ and $\psi$ fields defined as,
\bea
\chi &=& D_{DW}^{-\dagger}(m) {\cal P}   \phi \; ,\nn 
\psi &=& D^{-1}_{DW}(m)  D_{DW}(1) D^\dagger_{DW}(1) \chi \; . 
\eea
The extra cost of the second inversion might be acceptable due to
better scaling for large $L_s$ or be compensated by better inversion
algorithms for the first order system. Exploring these possibilities
is beyond the scope of this paper.

Finally, using the formalism described above we can formulate a
preconditioned HMC using a small $L_s$ domain wall operator as a
preconditioner. Then using a  multiple time scale integrator, one
can run most of the calculation using the small $L_s$ operator. The
correction step which will require the inversion of the large $L_s$
matrix will only be needed a few times during a trajectory. The
precise implementation as well as the possible interplay with the
Hasenbush preconditioner requires detailed experimentation for finding
the optimal HMC algorithm for DWF dynamical calculations. Here we
simply suggest one more trick to be used together with the rest of
today's HMC methodology.

%%%%%%%%%%%%%%%%%%%%%%%%%%%%  Section %%%%%%%%%%%%%%%%%%%%%%%%%%%%%%%%

\section{Red Black Preconditioning}
\label{sec:RedBlack}

In order to place the performance of M\"obius  on an equal
footing with Shamir Fermions, it is essential to implement red-black
preconditioning.  For the standard Shamir implementations, this step alone
accounts for nearly a factor of 3 speed up.  However the standard 5d 
even-odd preconditioning is not efficient for the M\"obius generalization because the
new Wilson operator $D_-$  connects 5d even and odd sites ($x \pm\mu$ and
$s \pm 1$).  This preconditioning would require a Wilson operator inverse as a
new inner loop. This problem  also occurs for the Bori\c{c}i domain wall
action and continued fraction approach to the Overlap operator and in general whenever the
hopping terms for the effective ``fifth'' dimension is coupled to the spatial
lattice.

To avoid this problem we define a new red-black partitioning that uses a 4d red-black lattice without alternating color as you run along the 5-th axis. The
result is a red-red and black-black matrix that is tridiagonal with constant
coefficients.  The specific construction is as follows.  Red-black is chosen
as checkerboard on the space-time lattice as $x_1 + x_2 + x_3 + x_4 =
\mbox{even/odd} $ for all $s \in [0,L_s] $.  The hopping matrix in 5d are
$D^{DW}_{br}$ and $D^{DW}_{rb}$ so
\be
D^{Wilson}(M) = 
\begin{bmatrix}
I_{rr} & D^{DW}_{rb} \cr
D^{DW}_{br} &  I_{bb} \cr 
\end{bmatrix}  \; ,
\ee
%{
with Schur decomposition, 
\be
D_{DW}(M) = 
\begin{bmatrix}
1 & 0 \cr
D^{DW}_{br}  I^{-1}_{rr} &  1 \cr
\end{bmatrix}
\begin{bmatrix}
I_{rr} & 0 \cr
0 &  I_{bb} - D^{DW}_{br} I^{-1}_{rr}  D^{DW}_{rb} \cr
\end{bmatrix}
\begin{bmatrix}
1 &  I^{-1}_{rr} D^{DW}_{rb} \cr
0 &  1  \cr
\end{bmatrix} \; ,
\ee
leads to the red-black Schur complement (or preconditioned matrix):
\be
D^{DW}_{pre} = 1 - I^{-1}_{bb} D^{DW}_{br}  I^{-1}_{rr} D^{DW}_{rb}  \; .
\ee
With $M_+ = b_5 (4 + M_5) + 1$ and $M_- = c_5 (4 + M_5)- 1 $,  the diagonal 
blocks in 5d are
\begin{eqnarray}
I_{rr} = I_{bb} = \begin{bmatrix}
M_+   & M_- P_- & 0 & \cdots &  - m P_+ \cr
M_- P_+ & M_+ & M_- P_- &\cdots  &  0 \cr
0 &  M_-  P_+ & M_+  &  \cdots   &  0 \cr
\vdots & \vdots & \vdots & \ddots   & \vdots \cr
- m  M_-  P_- & 0 & 0 &  \cdots  & M_+ \cr
\end{bmatrix}  \; .
\label{eq:D_5d_matrix}
\end{eqnarray}  
These can easily be inverted with the observation that
$I_{rr} = A_{rr} P_- + A^T_{rr} P_+$ where
\begin{eqnarray}
A_{rr} = M_+ \begin{bmatrix}
1   & M_-/M_+  & 0 &\cdots &   0  \cr
0     & 1 & M_-/M_+  & \cdots &  0 \cr
0 &   0     & 1  & \cdots  &  0 \cr
\vdots & \vdots & \vdots &  \ddots & \vdots \cr
- m M_-/M_+   & 0 & 0 &  \cdots    & 1  \cr
\end{bmatrix} \; .
\end{eqnarray}
To find the inverse $I^{-1}_{rr} = A^{-1}_{rr} P_- + A^{T -1}_{rr} P_+$, we
solve $A_{rr}\Psi_L = b_L$ and $A^\dagger_{rr} \Psi_R = b_R$ independently
for each chiral sector. This is performed by Gauss elimination of $O(L_s)$
steps of multiply-add and one division for half the lattice points per site in the
space-time volume.  There is no communication if the s-axis is inside the
processing element. The computation cost is negligible in comparison with
single Dirac applications.  

The closed form solution of $A_{rr} \Psi = b$ with $\alpha = M_-/M_+$ and $M = m M_-/M_+$
is first found for $s = L_s$ by forward eliminations,
\be
\Psi_{L_s} = \frac{1}{M_+ \; (1 -   M\;(-\alpha)^{L_s-1})}   [b_L + M\sum^{L_s -1}_{s=1} (-\alpha)^{s-1} b_{s}] \; ,
\ee
and the rest for $s = 1,...,L_s - 1$  by back substitution, 
\be
\Psi_s = - \alpha \Psi_{s+1} + \frac{b_s}{M_+} \; .
\ee
The same procedure interchanging rows and columns solves, $A^T_{rr} \Psi =
b$. However we emphasize that in the code it is best to do Gaussian
elimination on the fly rather than store the matrix inverse.

Our new (4d) red-black preconditioning for Shamir Fermions is
performing as well as the original 5d red-black preconditioning in
terms of iteration count. For M\"obius Fermions the speed up due to
our variant of red-black preconditioning is about a factor 2.5 for our
present lattice set. In fact the new version which treats all 5-th
axis sites uniformly at fixed x is advantageous in fast assembly level
code for vector pipelines such as those found in Intel processors.
Combining this with better data layout using Morton data layout, the
Shamir version of the M\"obius Domain Wall Fermion (MDWF) code of
Pochinsky~\cite{Pochinsky:2008zz} has achieved a factor 2 improvement
relative to earlier implementations.

%%%%%%%%%%%%%%%%%%%%%%%%%%%%  Section %%%%%%%%%%%%%%%%%%%%%%%%%%%%%%%%
\section{Numerical Results}
\label{sec:NumTests}

\begin{figure}[t]
\begin{center}
\includegraphics[width=0.7\textwidth]{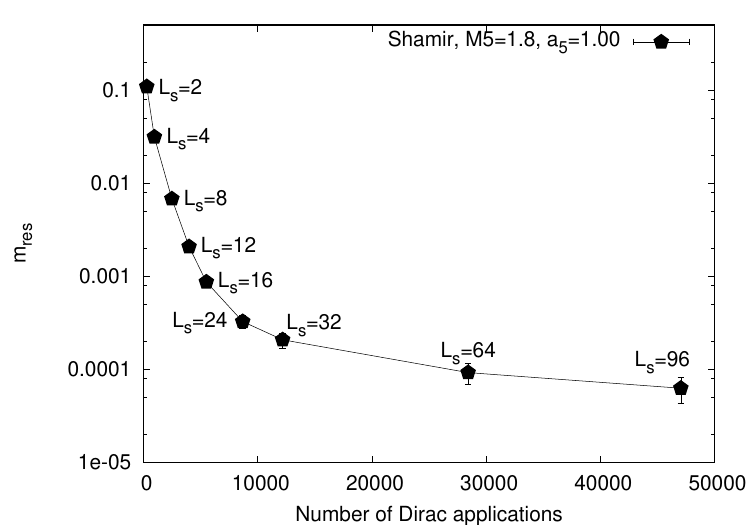}
\caption{$L_s$ dependence of the residual mass versus the number of
  Wilson Dirac applications required to invert the standard Shamir operator
  with $M_5 = 1.8$ and $a_5 = 1$ at $m = 0.06$. }
\label{Fig:fig0}
\end{center}
\end{figure}

We have made a series of tests of the effectiveness of
the M\"obius operator relative to  the standard Shamir formulation.
 For the  Shamir formulation (\ref{eq:Shamir}) where  $a_5$ is fixed to 1, one tunes
the Wilson mass $M_5$ and the extent of the fifth dimension $L_s$ in
order to achieve the desired degree of approximation to the exactly
chiral Dirac operator.  While the  violation of chirality  as measured by  $m_{res}$  appears to decrease exponentially at first 
for increasing $L_s$ , it  quickly slows down  asymptotically  to  an  $O(1/L_s)$ power fall off~\cite{Shamir:2000cf,Golterman:2005fe,Antonio:2008zz} as illustrated
in Fig.~\ref{Fig:fig0}.   

\begin{figure}[t]
\begin{center}
\includegraphics[width=0.7\textwidth]{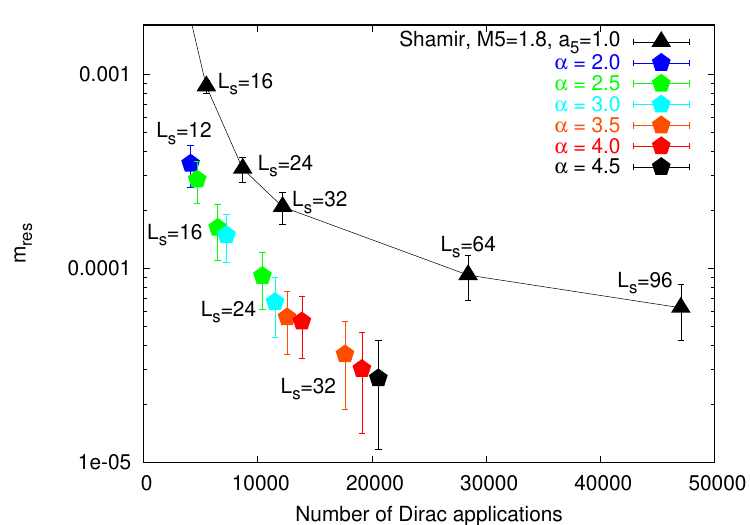}
\caption{ The residual mass for the M\"obius algorithm  as a
  function of $L_s$ and $\alpha$ at fixed  $M_5=1.5$,  $a_5=1$ and  $m=0.06$ in
  comparison with  the standard Shamir algorithm of Fig.~\ref{Fig:fig0}.}
\label{Fig:fig5}
\end{center}
\end{figure}

In the M\"obius kernel (\ref{eq:Moebius}) in addition to $M_5$ and
$L_s$, there are two more parameters, the lattice spacing in $5^{th}$
dimension, $a_5=b_5-c_5$, and the scaling parameter, $\alpha =
b_5+c_5$. The parameter $a_5$ represents a modification of $H_5$ and
allows us to interpolate between the Bori\c{c}i kernel and the Shamir
kernel. However the essential parameter in our scheme is the scaling
parameter, $\alpha$, which allows us to optimize the window (see
Fig.~\ref{Fig:fig7} ) for the sign function (\ref{eq:ApproxEPS}) at
fixed $L_s$. Before describing the details of our numerical
investigation, we present in Fig.~\ref{Fig:fig5} a summary of the
result of minimizing the chiral violation as measured by $m_{res}$ at
fixed $L_s$.
This clearly shows the importance of choosing an optimal value of
$\alpha(L_s) $ as a function of $L_s$. It is remarkable that the
residual mass now appears to continue to fall exponentially with
$L_s$, if we allow this re-tuning of $\alpha$ as $L_s$ increases.
However in Sec.~\ref{sec:violations} we argue that a tuned M\"obius
algorithm eventually falls asymptotically as $O(1/L^2_s)$, in contrast
to the standard Shamir action, which shows a much earlier cross over
from exponential to $O(1/L_s)$ power behavior at large $L_s$.

For very small $m_{res}$ the M\"{o}bius formulation has the potential
for orders of magnitude reduction of the explicit chiral symmetry
breaking at fixed computational cost. For example in our test case,
residual masses $6 \times 10^{-4}$ and $3 \times 10^{-5}$ were
achieved at $L_s=24$ and $L_s = 32$ respectively. Achieving the same
residual masses with the Shamir formulation would apparently require a
prohibitively large domain wall separations of order $L_s=10^2$ and
$L_s = 10^3$ respectively. This observation is consistent with the
estimate of an $O(1/L_s)$ vs $O(1/L_s^2)$ asymptotic fall off of
$m_{res}$ for the standard Shamir algorithm vs the improved M\"obius
algorithm.

\subsection{Testing procedure}

We have run a large number of tests but for simplicity we present here
an optimization study based only on a small sample of 20 quenched
$\beta=6.0$ Wilson gauge action configurations available from the
Gauge Connection archive. In spite of this small test sample, we
believe the lessons we have learned here
 are general based on our understanding of how the M\"{o}bius formulation works.

 Before we proceed in presenting the details of our testing procedure,
 we note that one M\"obius operator application does not need more
 Wilson Dirac operator applications than the standard Shamir operation
 at the same $L_s$. At first glance one might think that this is not
 the case due to the occurrence of the additional off diagonal terms
 (\ref{eq:MoebiusOp}) to $s \pm 1$ with the Wilson Dirac operator.
 However by first taking a vector gather of the neighboring Dirac
 spinors at $s+1,s,s-1$ no additional Dirac applications are required.
 The additional vector operation on the spinor is insignificant and
 can be ignored in the total cost estimates. In addition, in order to
 achieve convergence of the conjugate gradient solver for the M\"obius
 operator similar that achieved for the Shamir operator, the new 4d Red/Black
 preconditioning described in Sec.~\ref{sec:RedBlack} had to be
 invented. This preconditioning also adds an insignificant amount of
 extra computation. In all the graphs, the label {\it number of Dirac
   applications} represents the product of $L_s $ times the number of
 iterations per source to fixed precision.

In comparing the residual chiral symmetry breaking and the cost of the
quark propagator calculation between different values of the
parameters, we have at our disposal, we are faced with the fact that the
quark mass renormalization factor changes as one changes the kernel
operator $H_5$. Hence estimating the cost of the calculation at
constant physics, and the physical residual chiral symmetry breaking
is tricky.  In order to resolve this issue, we adopted the following
scheme in tuning our bare quark masses.  First, as a  point of
reference for all our tests we adopted the Shamir operator with
$M_5=1.8$ (with $a_5=1.0$ and
$\alpha=1.0$ in M\"obius parameterization). The RBC collaboration found
in~\cite{Blum:2000kn,Aoki:2002vt} that this $M_5$ is optimal for the
$\beta=6.0$ Wilson gauge action quenched ensemble. This was followed by a variation of the quark mass
on the M\"obius side, until the pion mass agreed with that obtained
using the standard Shamir action. The renormalization factors, $Z_m$, is the ratio of the Shamir bare quark mass to the M\"obius bare quark mass needed to obtain equal residual
masses.  
%As it turns out, the quark mass renormalization factor
%depends very little on $L_s$ and $\alpha$, hence no re-tuning of the
%bare quark mass was done when these two parameters were changed. 
%In Tables~\ref{tab:qmass} the dependence of the bare quark mass, keeping
%the pion mass fixed, on the parameters of the M\"obius action can be
%seen. 

From the data in Table~\ref{tab:qmass}, we can deduce that the quark mass
renormalization factor $Z_m$ is mostly sensitive to the parameter
$a_5 = b_5 - c_5$. All other parameters induce very small variations
on this renormalization factor. For that reason, in 
comparing residual masses between different sets of parameters, only
the non-trivial dependence of $Z_m$ on $a_5$ has to be taken into
account.
%We found that M\"obius with $L_s=8$, $a_5=1$ and $M_5=1.5$
%resulted in roughly the same residual mass as standard Shamir Fermions
%with $L_s=16$. 
In all our tests the Shamir quark mass we used is
$m=0.06$ which gave us a pion mass of roughly, $m_{\pi} = 0.44$ in
lattice units.
\begin{table}[thb]
%\begin{ruledtabular}
\begin{center}
\begin{tabular}{|cc|cc|cc|}
\hline
$m$ & $a_5$ & $m$ & $M_5$ & $m$ & $L_s$\\
\hline
0.130 & 2.00 & 0.093  &1.3  & 0.091& 12 \\
0.109 & 1.75 & 0.092 & 1.4 & 0.091& 16\\
0.091  & 1.50 & 0.091 & 1.5 & 0.091& 24 \\
0.075 & 1.25 &   0.090 &1.6 & 0.091& 32 \\
0.065 & 1.00 &  &  && \\
0.055  & 0.75 &  &  && \\
0.040 & 0.00 &  &  && \\
\hline
\end{tabular}
%\end{ruledtabular}
\caption{The quark mass, $m$, in the M\"obius algorithm at
fixed $m_\pi = 0.44 $. We vary 
$a_5$, $M_5$ and $L_s$, one at a time, away from the 
base case: $(a_5, M_5, L_s) = (1.5,1.5,16)$.
The dependence on $\alpha = b_5 + c_5$ is not visible
to the accuracy quoted above.}
\label{tab:qmass}
\end{center}
\end{table}

\subsection{Dependence on $a_5$ and $M_5$ at fixed $L_s$}

\begin{figure}[h]

\begin{center}
\includegraphics[width=0.7\textwidth]{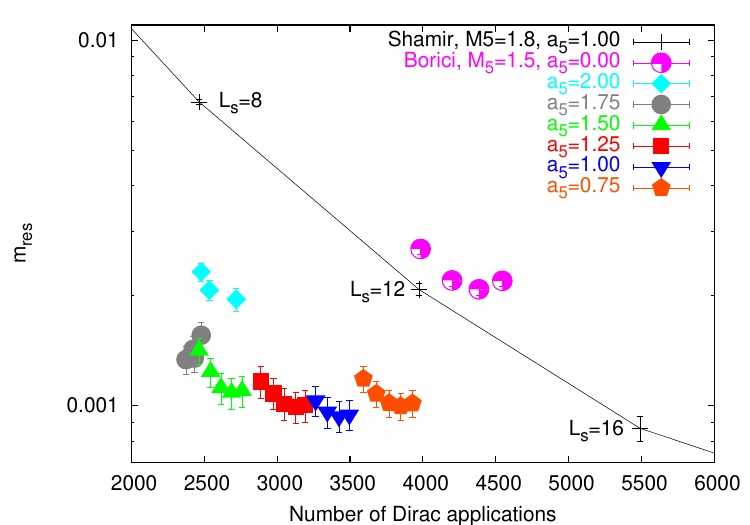}
\caption{The M\"obius algorithm  as a
  function of $a_5$ and $\alpha$ at fixed pseudoscalar mass,  $M_5=1.5$ and $L_s =8$ in
  comparison with the standard Shamir algorithm of Fig.~\ref{Fig:fig0}. The series of points for a given $a_5$ correspond to different values of
$\alpha$ enumerated in the text.}
\label{Fig:fig1}
\end{center}
\end{figure}

%\subsubsection{Results for quenched fields}

\begin{figure}[h]
\begin{center}
\includegraphics[width=0.7\textwidth]{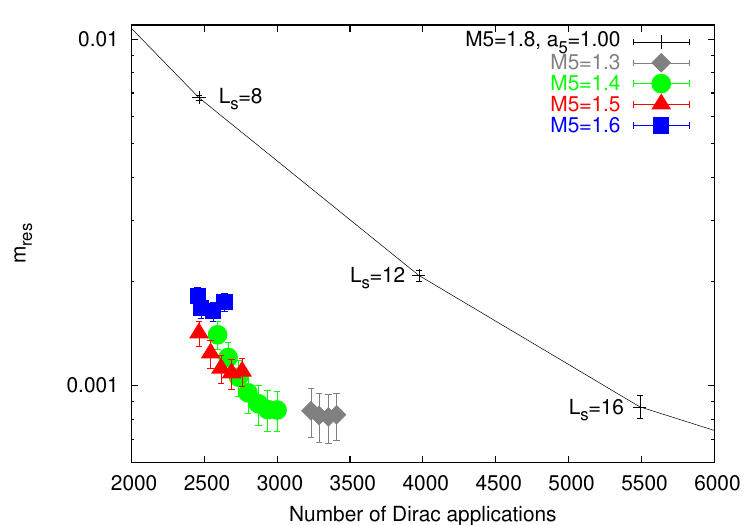}
\caption{ The M\"obius algorithm  as a
  function of $M_5$ and $\alpha$ at fixed pseudo scalar mass, $a_5=1.5$ and $L_s=8$ in
  comparison with the standard Shamir algorithm of Fig.~\ref{Fig:fig0}.
The series of points for a given $M_5$ correspond to different values of
$\alpha$ enumerated in the text. }
\label{Fig:fig3}
\end{center}
\end{figure}

An exhaustive exploration of the M\"obius parameter space $M_5, a_5, \alpha$ for minimizing
$m_{res}$ at fixed $L_s$ and $m_\pi$ is a laborious task. Fortunately it is easy to 
develop heuristics  that simplify the search. Here we describe two limited searches 
at $L_s =8$ to illustrate our procedure.  These explore the optimization of $\alpha$  for a range of values of $a_5$ at fixed  $M_5 = 1.5$ and subsequently we performed an optimization of  $M_5$ fixing $a_5$ to its optimal value $a_5 = 1.5$.

In Fig.\ref{Fig:fig1} we present the residual mass as defined in Eq.~\ref{eq:MRESCLASSIC}  for the M\"obius operator
at $L_s=8$ and $M_5=1.5$, varying $a_5$.  The series of
points for a given $a_5$ correspond to different values of
$\alpha$. In general, the number of Dirac applications grows with increasing $\alpha$. But
for $a_5=1.75$, it can be seen that this behavior changes and it is
even reversed at $a_5=2.0$.   From this graph we conclude that
 $a_5=1.5$ is the optimal point. It gives roughly the same residual mass
as the Shamir at $L_s=16$ with roughly half the Dirac operator applications.
To be precise  in  Fig.\ref{Fig:fig1}
the values of $\alpha$ are increased by $0.1$ 
moving left to right (for increasing Dirac applications)
in the intervals: $\alpha \in [1.0 - 1.3]$ for $a_5=0$, $\alpha \in [2.3-2.6]$ for
  $a_5=0.75$, $\alpha \in [2.2 - 2.5]$ for $a_5=1.0$, $\alpha \in [2.1- 2.4]$ for $a_5=1.25$
and  $\alpha \in [1.9-2.2]$ for $a_5=1.5$.   For  $a_5 = 2.0$ with
the values $\alpha = [1.6,1.9,2.0]$,   the points reverse direction moving right to left for decreasing Dirac application
and for $a_5=1.75$ in the cross over region    with  $\alpha = [1.7-2.1]$, the first three points move down to the left and last
point moves back to the right.
%even though $a_5=1.75$ shows an even better performance at
%$\alpha=1.9$ but turns out to be
%difficult to tune, due to the  sharp peak around this $\alpha$ value.
%For clarity we present  the same results in Fig.\ref{Fig:fig2}  plotting only
%the optimal $\alpha$ values for each $a_5$ value.

Next we fix $a_5 = 1.5$ to its optimal value and   explore the $M_5$ dependence of the residual mass
for M\"obius action shown in  Fig.~\ref{Fig:fig3}. Here as we increase $\alpha$ in all cases the number of Dirac applications increase.
In Fig.~\ref{Fig:fig3}  moving left to right (for increasing Dirac applications) with neighboring point separated by  $0.1$
we plotted   $\alpha \in [1.6 - 1.9]$ for $M_5=1.6$, $\alpha \in [1.9-2.2]$ for
  $M_5=1.5$, $\alpha \in [2.3 - 2.6]$ for $M_5=1.4$ and   $\alpha \in [2.6- 3.2]$ for $M_5=1.3$.
  It can be seen that $M_5=1.5$ and $M_5=1.4$ show equally good
  performance. $M_5=1.4$ achieves a little lower residual mass with a
  corresponding modest increase in the number of Dirac applications.
  In Fig.~\ref{Fig:fig5} summarizing our results for the optimal
  tuning, we choose $M_5 = 1.5$. Our overall conclusion is
  that the scaling parameter $\alpha$ is the most important for
  optimization. This trend is explained in more detail in our
  discussion of the residual mass in Sec.~\ref{sec:violations} below.
  We emphasize that if you choose to only minimize the residual mass
  by increasing $\alpha$ in accordance with the scaling rule in
  Eq.~\ref{eq:scaling} adjusting $\alpha \lambda_{max} < L_s $ for a
  fixed  M\"obius kernel, the spectral decomposition of the effective 4d
  operator is barely modified. So this procedure can
 be viewed as essentially as an improved algorithm for the same Dirac action, converging more rapidly
  to the same 4d overlap action in the  $L_s \rightarrow
  \infty$ limit.

Because we performed all our tests at fixed pion mass, one might ask
how this optimization depends on the bare quark mass. For that reason
we repeated our optimization of $\alpha$ at bare quark mass $m=
0.02$ and found that the choices we made in our original test at $m=
0.06$ are still optimal.

\subsection{Zolotarev Polynomials for the quenched fields}

\begin{figure}[t]
\begin{center}
\includegraphics[angle = 0,width=0.8\textwidth]{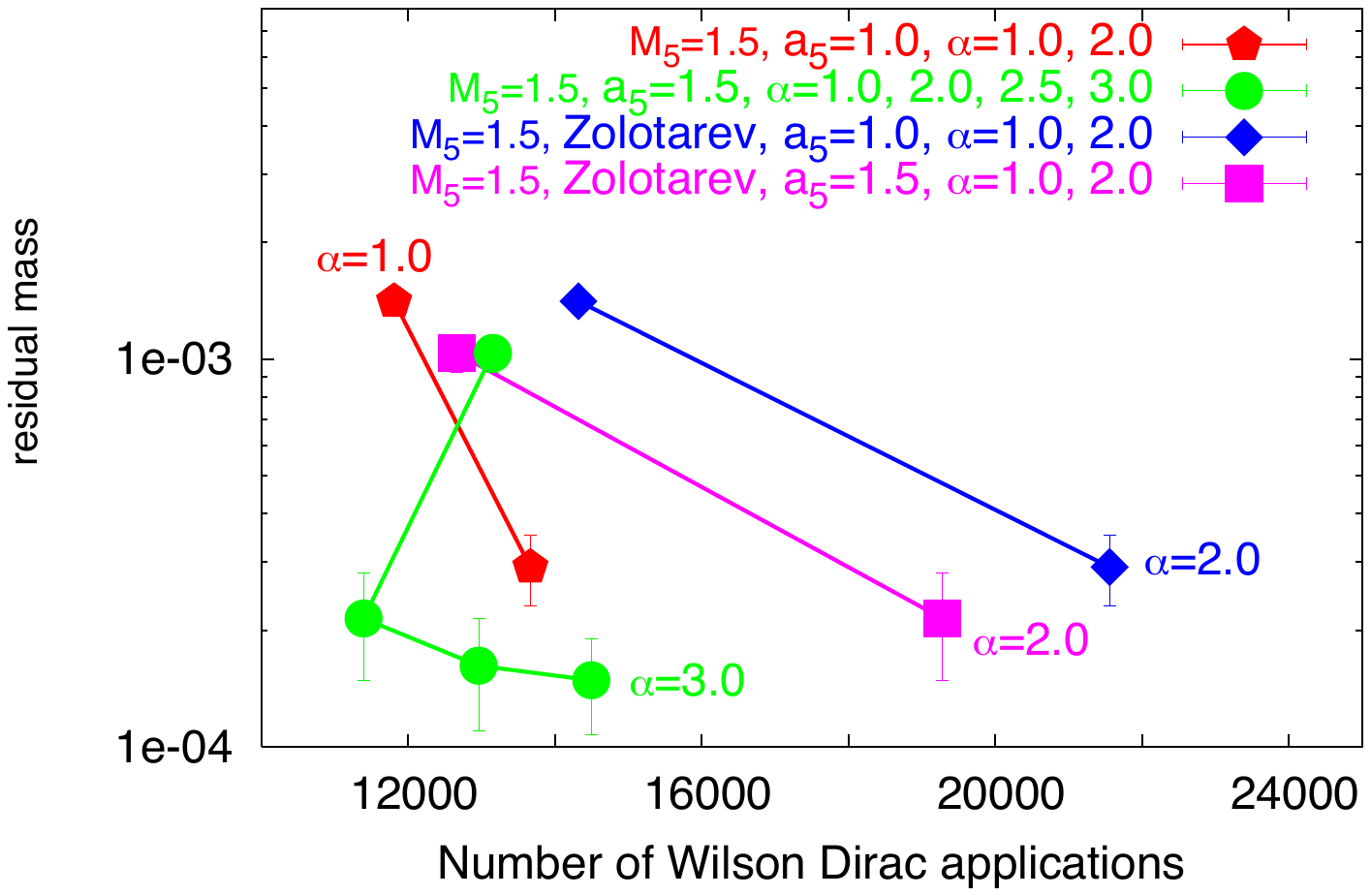}
\caption{Comparison of the Zolotarev and polar decomposition
  approximation to the sign function for varying $\alpha$ and two
  different $a_5$.}
\label{Fig:fig6}
\end{center}
\end{figure}

Finally, we should point out that the M\"obius formalism allows for
the use of the Zolotarev approximation to the sign
function~\cite{Chiu:2002ir,Chiu:2002kj} even for $a_5\ne 0$. One has
to introduce an $\alpha$ that is dependent on the fifth dimension. For
each 4d slice in the fifth dimension $\alpha_s$ has to be set to the
inverse of a root of the Zolotarev polynomial, i.e.\ for every $s \in
L_s$, $\alpha_s = b_s +c_s$ is equal to a Zolotarev coefficient, where
$b_s - c_s = a_5$. We define an $\alpha$ (without subscript), as an
overall factor, to act as $\alpha * \alpha_s$. In Fig.\ref{Fig:fig6}
we study the numerical behavior of the Zolotarev approximation, for a
setting as shown in Fig.\ref{Fig:fig7}, i.e.\ with $L_s = 10$ for
Zolotarev and $L_s=16$ for the polar decomposition, both positioned
logarithmically symmetric around one (i.e.\ per definition with
$\alpha = 1$).  Fig.\ref{Fig:fig6} shows results for $a_5 = 1.0$ and
$a_5 = 1.5$. For a given $\alpha$ the residual masses of the two cases
agree, as expected from the polynomials in Fig.\ref{Fig:fig7}. The
crucial observation is that for both values of  $a_5$, the Zolotarev
approximation, even though it employs a smaller $L_s$, required a
larger number of Wilson Dirac applications. So far we have not 
found any way to make Zolotarev competitive to the M\"obius in the  polar
form.

Furthermore, the reflection symmetry and the positivity of the
residual mass is lost by using the standard Zolotarev approximation.
One can restore these properties by using polynomials with double
roots and arranging them in pairs at locations $s$ and $L_s-s -1$ in
the fifth dimension. But the loss of positivity means that $m_{res}$
is no longer be an appropriate measure of chiral symmetry violations.  As
noted in the introduction a trivial shift in the bare quark mass by
$m_{res}$ only cancels the first term in the effective chiral
Lagrangian.  The higher order terms, such as the clover term, may or
may not be improved by the Zolotarev polynomial. Instead
for the Zolotarev approximation, a better measure of chiral
violation is to consider the  sign function,
$\Delta(\lambda) = (1-\epsilon^2(\lambda))/4$, which is uniform across the
approximation window $(\lambda_{min}, \lambda_{max})$. If
$\Delta(\lambda)$ becomes ${\cal O}(\varepsilon)$ for all $\lambda\in
(\lambda_{min}, \lambda_{max})$, then all chiral properties should be
also improved to that order.

\section{Domain Wall/Overlap Operator Correspondence}
\label{sec:equiv}

\begin{figure}[t]
\begin{center}
\setlength{\unitlength}{0.7 mm}
\begin{picture}(220,70)
\linethickness{.25mm}
\put(0,0){\large $y$}
\put(5,25){\large $D^{-1}_{ov}(m)$}  
%\put(0,10){\circle{2}}
\put(0,10){\line(2,1){50}}
\put(40,45){\large $J_\mu^{(a)ov}(x)$}  
\put(50,35){\circle{5}}
\put(100,0){\large $z$} 
\put(75,25){\large $D^{-1}_{ov}(m)$}  
\put(50,35){\line(2,-1){50}}
\put(110,40){\large $\equiv$}
\put(125,0){\large $y,1$} 
\put(130,25){\large $D^{-1}_{DW}(m) $}
\put(125,10){\line(2,1){50}} 
\put(155,45){\large $\sum_s j^{(a)DW}_\mu(x,s)$}  
\put(175,35){\circle{5}}
\put(225,0){\large $z,1$}  
\put(205,25){\large $D^{-1}_{DW}(m)$}
\put(175,35){\line(2,-1){50}}  
\end{picture}
\caption{The overlap current insertion in the quark propagator
is equivalent to a local 5d current insertion in the domain wall
bulk summed over the 5-th co-ordinate $s$.}
\label{Fig:X}
\end{center}
\end{figure}
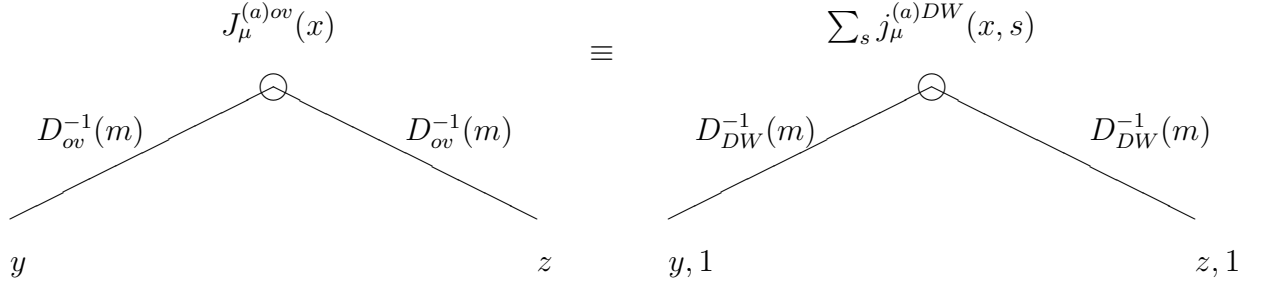

To complete the dictionary between overlap and its corresponding domain
implementation, one must show that all Fermionic correlators in a
fixed gauge background, $U_\mu(x)$, are equivalent. Namely that
\be
 \< {\cal O}[q,\overline q] \>_{DW}  =  \< {\cal O}[ \psi, \overline  \psi] \>_{ov} \: ,
\label{eq:corrId}
\ee
where here the domain wall correlators are restricted to Dirac fields on the boundaries,
(see Fig.~\ref{Fig:X}),
\be
 q_x = [{ \cal P}^\dagger \Psi]_{x,s=1}  \quad , \quad   {\overline q}_x  = [\overline \Psi D^{DW}(1){\cal P} ]_{x,s=1} \; .
 \label{eq:Qtilde}
\ee
The Fermionic path integrals $\< \cdots \>_{DW}$ and $\< \cdots \>_{ov}$  are given by
\be 
\< {\cal O}[ \psi, \overline  \psi] \>_{ov} =  \int d \overline \psi_{x} d\psi_{x} \; e^{\textstyle - \overline \psi_{x} D^{ov}_{x;y}(m)\psi_{y} } \; {\cal O}[ \psi, \overline  \psi]  \; ,
\label{eq:ovpath}
\ee
and 
\be
 \< {\cal O}[q,\overline q] \>_{DW}  = \int d \overline\Psi d\Psi d\overline \Phi d\Phi \;
e^{\textstyle  - \overline  \Psi_{x,s} D^{DW}_{x,s;,y,s}(m)\Psi_{y,s'}  - \overline  \Phi_{x,s} D^{DW}_{x,s;,y,s'}(1)\Phi_{y,s'} }  \; {\cal O}[q,\overline q]  \; , 
\label{eq:DWpath}
\ee
for overlap and  domain wall Fermions respectively.   
In the latter   $ \overline\Psi_{x,s}, \Psi_{x,s}$ are 5d Dirac fields and $\overline
\Phi_{x,s}, \Phi_{x,s}$ are 5d bosonic Pauli-Villars fields. 

Choosing the unit operator $O[q,\overline q] = {\bf1}$  the lowest order
identity  is 
\be
Z_{ov}[U] = Z_{DW}]U] \; ,
\ee
choosing the irrelevant field independent overall normalization constant appropriately.
This identity is a simple consequence of our  ``master'' equation (\ref{eq:Matrix}) for
$K^{DW} = {\cal P}^\dagger D^{-1}_{DW}(m) D(1)_{DW} {\cal P}$. 
It is also instructive to see this directly as a field redefinition,
$\Phi' = {\cal P^\dagger} \Phi \quad , \quad \overline{ \Phi'} =
\overline {\Phi} D^{DW}(1) {\cal P}$, in the domain wall partition
function. The result is that the Pauli-Villars action is trivial:
$\overline \Phi D^{DW}(1) \Phi = \overline \Phi' \Phi'$. To cancel the
Jacobian an identical redefinition is needed for the Dirac field, $Q =
{\cal P^\dagger} \Psi \; , \;\overline {Q} = \overline {\Psi}
D^{DW}(1) {\cal P} $. Thus one can rewrite the domain wall partition
function,
\bea
Z_{DW}[U] &=&  \int d \overline Q dQ  \;
e^{\textstyle  - \overline  Q {\cal P^\dagger} D^{DW}(1)^{-1}
  D^{DW}(m) {\cal P}Q } = det(K^{DW})  \; ,
\eea
which indicates again why the Dirac matrix, $K^{DW}$, preconditioned
by the Pauli-Villars matrix plays a central role. 

Using Wick's theorem on a general operator establishes the full set of identities. For example
Wick's theorem,
\be
 \< q_x \overline q_y \>_{DW} = 
[{ \cal P}^\dagger \underbracket{\Psi \ \overline \Psi} D^{DW}(1){\cal P} ]_{x1,y1} =
[{\cal P}^\dagger D^{-1 }_{DW}(m) D_{DW}(1) {\cal P}]_{x1,y1} = D^{ov -1}_{x,y}(m) \nonumber \; ,
\ee
gives the propagator identity,
\be
\< q_x \overline q_y \>_{DW} = D^{ov -1}_{x,y}(m)  = \< \psi_x \overline \psi_y \>_{ov} \; .
\ee
A formal derivation of all Fermionic correlators continues by iteration or
more systematically by introducing sources for the Fermion field as a generating function. 

In order to avoid confusion it is  important to note that  our {\bf anti-spinor}, 
\be
{\overline q}_x  = [\overline \Psi D^{DW}(1){\cal P} ]_{x,s=1} \,,
\ee
 in Eq.~\ref{eq:Qtilde} differs from the traditional choice~\footnote
{In particular  $\widetilde q = [\overline \Psi{\cal R} {\cal P}]_{s=1}$
for the Shamir action with $D_- = -1$. },
\be
\widetilde q_x = [\overline \Psi (- D_-){\cal R} {\cal P}]_{x,s=1} = 
[\overline \Psi (- D_-){\cal P}^\dagger]_{x,s=L_s} =  
(1-m)^{-1}[\overline \Psi (D_{DW}(1) - D_{DW}(m)){\cal P}]_{x,s=1} \;, 
\ee
or $\overline q_x = (1-m) \widetilde q_x + [\overline \Psi D_{DW}(m) {\cal
  P}]_{x,s=1}$.
The traditional  definition results in  the awkward subtracted overlap quark propagator
\be
 \< q_x \widetilde q_y \>_{DW} = \frac{1}{1-m}[D^{-1}_{ov}(m) - 1]_{xy} \; .
\ee
 Our new definition of $\bar q$ smears the boundary field by
one lattice unit by virtue of the application of  $D^{DW}(1)$ operator. We believe this is a better choice since it
simplifies the identities between domain wall and overlap
correlation functions, suppressing reference to the
particular implementation be it Shamir, M\"obius or some
future variant.  Moreover, an advantage of our definition  of
$\overline q_x$ is that changing the normalization of the operator
$D_{DW}(m)$ has {\bf no} effect on the correlation functions since
this change cancels with $D^{-1}_{DW}(1)$. 

For our discussion of the axial current, it is
useful to extend into the bulk both definitions of the anti-spinor, $\overline Q_s =
[\bar\Psi D^{DW}(1) {\cal P}]_s$ and $\widetilde Q_s = [\bar\Psi (- D_-) {\cal
  P}^\dagger]_s$ for all s with their respective correlation
matrices. These definitions give the following bulk to bulk 5d correlators
\be
A^{DW}_{ss'}(m) = \<Q_s \overline  Q_{s'}\> = [{\cal P}^\dagger \frac{1}{D^{DW}(m)}D^{DW}(1){\cal P}]_{s,s'}
\label{eq:Amatrix}\,,
\ee
given in Eq.~\ref{eq:DWFmatrix} above, and 
\be
M_{s,s'}(m) = \<Q_s \widetilde Q_{s'}\> = [{\cal
    P}^\dagger \frac{1}{D^{DW}(m)}( -D_-){\cal P^\dagger}]_{s,s'}
\label{eq:Matrix}\,,
\ee
respectively.  The evaluation of these matrix elements (or correlators) are given in Appendix ~\ref{sec:A}. The mass dependence  enjoys the nice factorization property, $M^{DW}(m) = A^{DW}(m) M^{DW}(1)$.

%%%%%%%%%%%%%%%%%%%%%%%%%%%%  Section %%%%%%%%%%%%%%%%%%%%%%%%%%%%%%%%

\subsection{Domain Wall/Overlap correspondence for currents}
%\newpage
%\section{Vector and Axial Ward-Takahashi Identities }
%\section{Domain Wall/Overlap correspondence for currents}
\label{sec:currents}

In the continuum the vector and axial current can be defined by Noether's
theorem through the local change of variable in the path integral,
\bea
\psi(x) &\rightarrow& \exp[i \theta^a_V(x) \lambda^a + i \theta^a_A(x) \lambda^a\gamma_5] \; \psi(x) \nn
\overline \psi(x) &\rightarrow& \overline \psi(x) \; 
\exp[-i \theta^a_V(x) \lambda^a + i \theta^a_A(x) \lambda^a \gamma_5] \,,
\eea
resulting in Ward identities for the divergence of local currents.
The current is then identified after integration by parts.  An
equivalent method is to {\em gauge} the action with a flavor gauge
field and define the current as the linear response to this gauge
field.  On the lattice the gauge approach is superior
particularly for non-local actions such as the overlap action. For
non-local actions, the analog of integration by parts (i.e. {\em summation by parts}'),
needed to identify the current in Noether's approach,  is difficult to
define.

For both the singlet and non-singlet vector current the ``gauging'' of
the action on the lattice action is accomplished by the substitution
$U_\mu(x) \rightarrow \exp[i \lambda^a A^a_\mu(x)] U_\mu(x)$ on
each link, where $A^a_\mu,(x)$ is an adjoint flavor gauge field on
link $(x,x+\mu)$.  Now by applying the {\bf Domain Wall/Overlap equivalence}, we must have
an  equivalence between the matrix element for vector  and axial currents.
From 
\be
-i \delta_{A^a_\mu(x)} \< \psi^i_y \overline\psi^j_z \>_{ov} = 
-i \delta_{A^a_\mu(x)} \< q^i_y \overline q^j_z \>_{DW} \,,
\ee
 we obtain the elegant identity,
\be
\< J_\mu^{(a)ov}(x) \; \psi^i_y \overline\psi^j_z \>_{ov} = \< J_\mu^{(a)DW}(x) \; q^i_y \overline q^j_z \>_{DW} \; ,
\label{eq:Vcorr}
\ee
neglecting for the moment a contact term that only contributes at   $x = z$.
The quark flavor indices are labeled by $i,j = 1, \cdots, n_f$.
The vector current operators resulting from variations of the
action are
\be
J^{(a) ov }_\mu(x) = -i  \delta_{A^a_\mu(x)}\overline  \psi_y D_{ov}(m)_{y,z}\psi_z =
 \overline \psi \lambda^a V^\mu(x) \psi \,,
\ee
and 
\be
J^{(a) DW }_\mu (x)  =  -i \delta_{A^a_\mu(x)} \overline \Psi  D_{DW} \Psi  + \cdots
 = \overline \Psi \lambda^a {\cal V}_\mu(x) \Psi  + \cdots \,,
\label{eq:DWvectorCurrent}
\ee
for the overlap and domain wall actions respectively.    To derive this is straight
forward resulting in 
\bea
 [D^{-1}_{ov}(m) (\delta_{A^a_\mu(x)} D_{ov}(m)  )
 D^{-1}_{ov}(m)]_{yz} &=& \<q_{y} \overline \Psi\>_{DW} \;
 (\delta_{A^a_\mu(x)}  D_{DW} )\<\Psi \overline q_{z} \>_{DW}  \nonumber \\
& -& \< Q_y \; \delta_{A^a_\mu(x)} \overline q_z \>_{DW}  \; .
\label{eq:VecID}
\eea 
The last term on the RHS is the contact term due to the
dependence of $\overline q = [\Psi D^{DW}(m) {\cal P}]_1$ on the gauge
fields. It only contributes to $x = z$, so it can be ignored
except at co-incident points. In the Shamir case the local
source $\widetilde q = [\Psi {\cal P}^\dagger]_1$ can be used to avoid
this contact term but as described above it then introduces into the
propagator a contact term and it is awkward to generalize this to M\"obius
Fermions. Again as a consistency check, it is worth noting that given
the bulk to boundary propagators $\<q \overline \Psi_{s'}\>_{DW} =
[{\cal
  P}^\dagger D^{ -1}_{DW}(1)]_{1s'}$ and $\<\Psi_s \overline q \>_{DW} =
[D^{ -1}_{DW}(m) D_{DW}(1) {\cal P} ]_{s1}$,  discussed in the Appendix~\ref{sec:A} , this identity (\ref{eq:VecID}) 
is equivalent to the variational derivative  of our fundamental equation (\ref{eq:inverseDWFmatrix}),
\be
 \delta_{A^a_\mu(x)}  D^{-1}_{ov}(m) = \delta_{A^a_\mu(x)}   [{\cal P}^\dagger \frac{1}{D_{DW}(m)} D_{DW}(1) {\cal P}]_{11},
\ee
including the contact term.  This leads to an explicit vector current overlap kernel  as a sum over the bulk domain wall modes,
\be
V^\mu_{yz}(x)  = D^{ov}_{yy'}(m) \<q_{y'} \overline \Psi_{y'',s'}\>_{DW} {\cal V}^\mu(x)_{y''s',x's} \<\Psi_{x',s} \overline q_{z'} \>_{DW}  D^{ov}_{z'z}(m)  \; .
\ee
Like the overlap propagator itself, this current  is  non-local in 4d falling off
exponential in $|y-x|$ and $|z-x|$. The $+ \cdots$ in Eq.~\ref{eq:DWvectorCurrent} refer to the Pauli-Villars term
which does not   contribute  to the  Fermionic matrix element being consider here. 
 However, as we will show later, there are occasions such as conserved current-current correlators, where the
Pauli-Villars contribution is {\bf required}. It follows from the identity,
\be
\< J^{(a) ov}_\mu(x) J^{(b) ov}_\nu(y) \> = \delta_{A^a_\mu(x)} \delta_{A^b_\nu(y)} \log[Z_{ov}] \equiv \delta_{A^a_\mu(x)} \delta_{A^b_\nu(y)} \log[Z_{DW}]\,,
\ee
that the correlator is 
\be
\< J^{(a) ov}_\mu(x) J^{(b) ov}_\nu(y) \> 
 = \sum_s \< j^{a DW}_\mu(x,s) j^{b DW}_\nu(y,s) + j^{a PV}_\mu(x,s) j^{b PV}_\nu(y,s)\> \; ,
\ee
as explained in detail in Sec.~\ref{sec:WTid}.

Defining the axial current directly from the overlap action is more
difficult because the global axial symmetry, which is realized by
$\widehat \gamma_5 = \gamma_5(1 -D_{ov}(0))$, is non-local and depends
on the background gauge field, $U_\mu(x)$. Thus for the axial current
we choose to {\bf define} the overlap axial current by imposing the
condition,
\be 
 \< J^{5 (a) ov}_\mu(x) \; \psi^i_y \overline\psi^j_z\>_{ov} = \< J^{5(a)DW}_\mu(x) \; q^i_y \overline q^j_z \>_{DW} \; . 
\label{eq:Acorr}
\ee

\section{Ward-Takahashi Identities}
\label{sec:WTid}

We now proceed to give explicit expressions for the vector and axial current
for the M\"obius domain wall action. We follow closely the literature
in particular the paper of Furman and Shamir~\cite{Furman:1995ky}.
Since the 5d domain action is a function of the 4d Wilson kernel, we
begin by reviewing the Wilson vector current to establish notation and
useful intermediate kernels. This vector kernel contributes to both
the vector and axial current.

\paragraph{Wilson Vector Current:} The vector current for Wilson Fermions
is well know but let us repeat the argument to establish the notation and methodology for the
chiral Fermions.  By gauging the Fermions action (\ref{eq:D^{Wilson}}) ,
\be
S^{Wilson}[U_\mu(x)] = \overline \Psi D^{Wilson}[U_\mu(x),M_5]  \Psi \,,
\ee
we compute the first order variation to get the vector current,
\be
{\mathbb J}^a_\mu(x) =  -i \delta_{A^a_\mu(x)} S^{Wilson}[U_\mu(x)e^{i \lambda^a
  A^a_\mu(x)} ]_{A^a_\mu(x) =0} \equiv \overline \Psi_z{\mathbb V}^\mu_{zy}(x) \lambda^a
\Psi_y\,,
\ee
where the ultra local kernel is
\be
{\mathbb V}^\mu_{zy}(x) = \frac{\gamma_\mu-1}{2} U_\mu(x) \delta_{z,x} \delta_{x+\mu,y}  + \frac{\gamma_\mu+1}{2} U^\dagger_\mu(x) \delta_{z,x+\mu} \delta_{x,y}\; .
\label{eq:Wkernel} 
\ee

Next we consider the  linear responses to change of integration variables, 
 $\Psi_x
\rightarrow \exp[i\lambda^a \theta^a_x] \Psi_x$, $ \overline \Psi_x \rightarrow
\overline \Psi_x \exp[i\lambda^a \theta^a_x]$ in the path integral and applying Neother's
theorem to get a  divergence condition of this same current. The change in the action is

\bea
-i \delta_{\theta^a_x} S^{Wilson}  &=& 
\overline \Psi_x \frac{\gamma_\mu-1}{2}\lambda^a U_\mu(x) \Psi_{x+\mu}  +  \overline \Psi_{x+\mu} \frac{\gamma_\mu+1}{2} \lambda^a U^\dagger_\mu(x) \Psi_x
- (x \rightarrow x-\mu) \nn
&=& {\mathbb J}^a_\mu(x) -  {\mathbb J}^a_\mu(x-\mu)  \equiv \Delta_{-\mu} {\mathbb J}^a_\mu(x)  \,.
\eea
By  Noether theorem we have  the lattice current conservation condition,~\footnote{ Of a course ``local''  currents like gauge variable $U_\mu(x)$, on the lattice, $J_\mu(x) \equiv J(x,x+\mu)$, are really bilocal variables
assigned to a positive links, $(x,x+ \mu)$.    The current in the negative direction carries the opposite sign: $ J(x +\mu,x)  \equiv - J_\mu(x)$.}
\be
\sum_\mu {\mathbb J}^a_\mu(x) -  {\mathbb J}^a_\mu(x-\mu)  \equiv \overline \Psi_x \lambda^a \frac{\partial L}{ \partial \overline \Psi_x} -  \frac{\partial L}{ \partial  \Psi_x}\lambda^a \Psi_x  = 0 \, ,
\ee
using the classical equations of motion.     Actually the divergence condition or Ward-Takahashi identity is a quantum
constraint inside the  path integral,
\be
\< \Delta_{-\mu} J^a_\mu(x) \;  {\cal O} \> =  \< \delta_{\theta^a_x}{\cal O}
\> \; , 
\ee
which is often given the short hand notation by asserting that,
$\Delta_{-\mu} J^a_\mu(x) = 0$, is an ``operator'' equation.  If the
action or the measure is not invariant additional terms need to be
included.

\subsection{Domain Wall Vector Current} 
\label{sec:DWvector}

The derivation of the vector current for the
domain wall action parallels the Wilson case closely.  Again we gauge the action and take the variation,
\be 
J^{a DW}_\mu(x)  = -i \delta_{A^a_\mu(x)}  \overline \Psi_{xs} D_{DW}(m)_{ys,zs'}[U_\mu] \Psi_{zs'}    -i \delta_{A^a_\mu(x)}\overline \Phi_{xs} D_{DW}(1)_{ys,zs'}[U_\mu] \Phi_{zs'} \; .
\ee
 For the Shamir action the
result is 
\be
 J^{a DW}_\mu(x) =  \sum^{L_s}_{s=1}[ \overline \Psi_{s} \lambda^a  {\mathbb   V}^\mu(x) \Psi_s + \overline \Phi_{s} \lambda^a  {\mathbb   V}^\mu(x) 
\Phi_{s}]\,,
\ee
expressed as the Wilson current averaged over all ``flavors'' in the
5-th direction.  The second term involves the Pauli-Villars fields, which
 give zero contribution to quark correlators. The generalization to the M\"obius vector current is straight forward.
Taking the variation of the domain wall action in Eq.~\ref{eq:DWaction} we obtain
\bea
 J^{a DW}_\mu(x)  &=& b_5 \sum^{L_s}_{s=1} \overline\Psi_s \lambda^a  {\mathbb   V}^\mu(x) \Psi_s 
      +      c_5  \sum^{L_s}_{s=2} \overline  \Psi_s  \lambda^a  {\mathbb   V}^\mu(x) { P_+ }\Psi_{s-1} 
      +       c_5  \sum^{L_s-1}_{s=1}   \overline \Psi_s  \lambda^a  {\mathbb   V}^\mu(x) {  P_- }  \Psi_{s+1}  \nonumber \\
      &- &   c_5  m \left(\overline \Psi_1 \lambda^a  {\mathbb   V}^\mu(x)  P_+\Psi_{L_s} {\bf +}  \;  \overline \Psi_{L_s} \lambda^a  {\mathbb   V}^\mu(x) P_-\Psi_{1} \right) + (\Psi, \overline \Psi, m \rightarrow \Phi,\overline \Phi,  1) \; .
\label{eq:DWvector}
\eea
The Pauli-Villars contribution is given exactly by the substitution
indicated above. It may seem surprising that the domain wall vector
current operator for M\"obius Fermions have explicit bare mass
dependence but the identity (\ref{eq:Vcorr}) with the overlap form
requires this. Indeed, in view of the form of the overlap action, $
S_F =\overline (1-m) \psi [ m/(1-m) + D_{ov}(0) ]\psi$ we see that the
effective overlap vector current is independent mass except for the
overall factor of $(1-m)$ which as we mentioned before could be
corrected trivially by renormalizing the field by $\sqrt{1-m}$ and
thus removing all mass dependence from the vector current.
 
 An alternative approach to the 4d vector current is to begin with constructing a local 5d conserved vector current, 
\be
\Delta_{-\mu} j^{a DW}_\mu(x,s) + \Delta_{-\hat 5} j^{a DW}_5(x,s) = 0 \; .
\ee
 This is accomplished in the same manner as before except now the domain wall action is gauged by 
a flavor potential, $A^a_\mu(x,s)$, depending on  the fifth dimension $s$.  The 4d current in Eq.~\ref{eq:DWvector}.
  is formed by summing over the 5-th axis
\be
 J^{a DW}_\mu(x) = \sum_s  j^{a DW}_\mu(x,s) \,,
\ee
and since no current leaks out of the 5-th axis,
\be
\sum_s \Delta_{-\hat 5} j^{a DW}_5(x,s) = 0 \; ,
\ee
it is conserved as a   4d vector current.

To get the full expression for the 5d vector for M\"obius Fermions,  we need to
take care of  off-diagonal terms $D_- P_{\pm}$ by gauging, 
\be
D_-[U_\mu(x)] \; P_\mp \rightarrow 
D_-[U_\mu(x) e^{\textstyle i \lambda^a A^a_\mu(x,s)}] \; P_\mp e^{\textstyle \pm i \lambda^a A^a_{\hat 5}(x,s)} \; .
\ee 
The resulting   5d ``vector'' current $j^a_M(x,s) = j^{a DW}_M(x,s) + j^{a PV}_M(x,s)$ is 
\bea
j^{a DW}_\mu(x,s) &=& b_5 \overline \Psi_{s} \lambda^a {\mathbb V}^\mu(x)  \Psi_{s} 
+  c_5 \overline \Psi_{s} \lambda^a {\mathbb V}^\mu(x) P_- \Psi_{s+1}
+  c_5 \overline \Psi_{s} \lambda^a {\mathbb V}^\mu(x) P_+ \Psi_{s-1} + \nonumber \\
&-& c_5 (1+m) \overline \Psi_{1} \lambda^a {\mathbb V}^\mu(x) P_+ \Psi_{L_s} \;\ \delta_{s,1}  + \nonumber \\
&-& c_5 (1+ m)\overline \Psi_{L_s} \lambda^a {\mathbb V}^\mu(x)  P_- \Psi_{1}  \;  \delta_{s,L_s} 
\eea
and 
\bea
j^{a DW}_{\hat 5}(x,s)& =& 
 \overline\Psi_{x,s} \lambda^a D_- P_- \Psi_{x,s+1} 
 - \overline \Psi_{x,s+1} \lambda^a  D_- P_+ \Psi_{x,s}  + \nonumber \\
&-&(1+ m) \overline \Psi_{x,L_s} \lambda^a  D_- P_- \Psi_{x,1} \; \delta_{s,L_s} + \nonumber \\
&+& (1+m) \overline \Psi_{x,1} \lambda^a D_- P_+ \Psi_{x,L_s}  \; \delta_{s,L_s} \;\;\;.
\eea
where $s+1$ is considered modulo $L_s$. 
Again the Pauli-Villars current $j^{a PV}_M(x,s)$ takes the same form with
substitution $\Psi, \overline \Psi \rightarrow \Phi, \overline \Phi$ and $m
\rightarrow 1$.

\subsection{Domain Wall Axial current} 

The subtlety for the axial current for the domain wall action is that
in 5d there is no chirality. The only local current is the 5d
vector current defined above. The analog for 4d chirality is the
parity transformation that reflects the 5-th axis exchanging the two
domain walls. This is the essence of the descent
relations~\cite{Callan:1984sa} that inspired Kaplan's use of the
domain construction in the first place.

To implement this on the lattice, one simply splits the extra
dimension on any link $(M,M+1)$ between the two domain walls ($1 < M <
L_s - 1]$) and defines local vector currents on each side. By gauging
each side separately we get a Left current for $s \in [1,M]$ and
Right current for $s \in [M+1,L_s]$. The 4d Ward identities follows
from Gauss's law applied to an open interval $s \in [s_0, s_1]$ where
the outgoing links $(s_0-1,s_0)$ and $(s_1,s_1+1)$ on the boundary are
dropped -- parallel Dirichlet branes if you like. Now flux
conservation requires including the flux through these boundaries,
\be
\sum^{s_1}_{s=s_0} \Delta_{-\mu} j^{a}_\mu(x,s) =   j^{a}_5(x,s_0-1) - j^a_5(x,s_1) \; .
\ee
Although the definition of the axial current can be done using an arbitrary point in the $5^{th}$ dimension to split the left handed current from the right handed current, we will restrict ourselves to the case where $M$ is taken to be $L_s/2$. Hence, we define the axial (and vector) currents by separating the left and right terms,
\be
J^\mu_L(x) = \sum^{L_s/2}_{s=1} j^{a }_\mu(x,s)  \quad \mbox{and} \quad
J^\mu_R(x) = \sum^{L_s}_{s=L_s/2+1} j^{a}_\mu(x,s) \; .
\ee
With periodic boundary condition $J^{a }_5(x,L_s) - J^{a}_5(x,0) = 0$, we
have the conserved vector current,
\be
J^{a DW}_\mu(x) \equiv  J^\mu_L(x) + J^\mu_R(x) = \sum^{L_s}_{s=1}j^{a}_\mu(x,s) \; .
\ee
The odd parity axial current is found by subtraction,
\be
 J^{(5)\mu}_{DW}(x) = J^\mu_L(x) - J^\mu_R(x) = \sum^{L_s}_{s=1}
 \Gamma_5(s-L_s/2) j^{a}_\mu(x,s) \;,
\ee
where $\Gamma_5(s-L_s/2) = \epsilon(L_s/2+1/2 -s)$. By Gauss' law we obtain the 
Ward-Takahashi identity,
\be
\Delta_{-\mu} J^{(5)\mu}_{DW}(x) = -2 j^a_{\hat 5}(x,L_s) + 2 j^a_{\hat 5}(x,L_s/2)\,,
\ee
where
\bea
   j^a_{\hat 5}(x,L_s) &=& - m \, \widetilde q_x \lambda^a \gamma_5 q_x  -  \widetilde \Phi_{x,L_s} \gamma_5 \lambda^a \Phi_{x,1}  \nn
 j^a_{\hat 5}(x,L_s/2)  &=&   \widetilde Q_{x,L_s/2} \gamma_5 \lambda^a Q_{x,L_s/2+1} +  \widetilde \Phi_{x,L_s/2} \gamma_5 \lambda^a\Phi_{x,L_s/2+1}   \; .
\eea 

\subsection{Overlap Axial Current}

The remaining task is to use the map between domain wall to overlap Fermions,
\be 
 \< J^{5 (a) ov}_\mu(x) \; \psi^i_y \overline\psi^j_z\>_{ov} = \< J^{5(a)DW}_\mu(x) \; q^i_y \overline q^j_z \>_{DW}  + i \< Q^i_y \; \delta_{A^a_\mu(x)} \overline q^j_z \>_{DW}  \; ,
\label{eq:AxialMap}
\ee
to express the divergence in terms of overlap fields.  Evaluating
\be
 \< \Delta_{-\mu} J^{(5)\mu}_{ov}(x) \; \psi_y
\overline\psi_z \>_{ov} \;   =  \<  (2 m \, \widetilde q_x \lambda^a \gamma_5 q_x 
+ 2 \widetilde Q_{x,s} \gamma_5 \lambda^a Q_{x,s+1})  \; q_y \overline q_z \>_{DW}\,,
\label{eq:DIV}
\ee
where $s=L_s/2$, implies,
\be
\Delta_{-\mu} J^{(5)\mu}_{ov}(x) =  m \overline \psi_x[(\gamma_5 + \widehat
\gamma_5) \psi]_x + 2 (1-m) \overline \psi_z \gamma_5 \rho^{zy}_{L_s}(x) \psi_y\,,
\ee
as we may readily check. Using the identity:  $2 \gamma_5 = (1-m) ( 1 + \gamma_5 \widehat \gamma_5) + 2 D_{ov}(m)$
the  first term in Eq.~\ref{eq:DIV}  gives,
\be
\frac{2m}{(1-m)} [ D^{-1}_{ov}(m) - 1]_{yx}  \gamma_5 [D^{-1}_{ov}(m)]_{xz}  
= m [D^{-1}_{ov}(m)]_{yx} [(\gamma_5 + \widehat \gamma_5) D^{-1}_{ov}(m)]_{xz}  + \mbox{CT} \; ,
\ee
where the contact terms (CT) are easily identified as 
\be
\mbox{CT} = -   \frac{2m}{(1-m)}  [ \delta_{yx} \gamma_5 D^{-1}_{ov}(m)_{xz} -     D^{-1}_{ov}(m)_{yx} \gamma_5 \delta_{x,z}]  \; .
\ee
One may absorb these terms in an appropriate introduction  of contact terms  in the 
map from domain wall to axial currents (Eq.~\ref{eq:AxialMap}) analogous
to those found for  the vector current but the 
procedure is rather arbitrary and unphysical.  The second term in Eq.~\ref{eq:DIV} is
\be
 \< q \widetilde Q_{x,s+1}\>\gamma_5 \<Q_{x,s}\overline q \> = (1-m) D^{-1}_{ov}(m)\gamma_5 \Delta^L_{s+1}  \Delta^R_{s}  D^{-1}_{ov}(m) \,,
\ee
where we define the Left and Right breaking term by
\be
 \Delta^L_{zx}  \Delta^R_{xy}  =   [\frac{T^{-1}_{L_s}...T^{-1}_{L_s/2+1}}{1 +  {\mathbb T}^{-L_s}}]^{zx}  
[\frac{T^{-1}_{L_s/2} ...T^{-1}_{1} ..}{1 +  {\mathbb T}^{-L_s}}]^{xy} 
\equiv \rho^{zy}_{L_s}(x)\,,
\label{eq:LRdelta}
\ee
whose space-time average $\sum_x \rho_{L_s}(x) = \Delta_{L_s}$ is
precisely the correct breaking term for global chiral symmetry found
earlier for the overlap action~(\ref{eq:global}). In addition note
that due to the conservation of 5d flux, the result is {\bf independent}
of the location ($s=L_s/2$) of the mid plane slice. Changing the position
merely redefines the axial current by a term with zero total
divergence. Nonetheless the natural definition is to take the mid-plane
so that parity is equivalent to the reflection operator, ${\cal R}: s
\rightarrow L_s -s$.

So far we have dealt with the non-singlet sector for the currents.
This singlet sector requires some special considerations. In the
vector channel one wishes to introduce finite chemical potential for the
Baryon number. In the axial channel one needs to show how the axial
anomaly arises in the domain wall formalism. In both cases a natural
representation in our domain wall/overlap correspondence exists.

\subsection{ Axial anomaly}
\label{sec:AxialAnomaly}

The flavor singlet domain wall axial current is anomalous, as has been
shown by Kikukawa and Nuguchi~\cite{Kikukawa:1999sy}. The way this
comes about is instructive.  This has been computed from the
self-contraction of the Fermionic fields,  $Q$ and $\widetilde Q$, in the bilinear term $\widetilde  Q_{L_s/2}(x)\gamma_5
Q_{L_s/2+1}(x)$  of Eq.~\ref{eq:DIV} at the
mid-plane of the fifth dimension. This term gives rise both to a contribution to $m_{res}$ (through contraction with the boundary fields)  and to a new
term that survives in the $L_s\rightarrow \infty$ limit. 
In order to rigorously perform the calculation, one needs to use the  map between domain wall and overlap
currents that requires both the domain wall Fermions and the Pauli-Villars
``bosons". The result is that this Fermion contribution at the mid-plane is
exactly canceled by the Pauli-Villars contraction and replaced by a
boundary contribution at the domain wall. This is really the correct
way to understand the physics. For example suppose we modified the
domain wall implementation as for example suggest in Ref.~\cite{Bar:2007ew}
by allowing slightly non-uniform gluon fields, $U_\mu(x,s)$, as a
function of the fifth coordinate. Without the Pauli-Villars
cancellation, the chiral quarks at the boundary would ``feel'' the
wrong gauge potential at the mid-plane giving unphysical and indeed incorrect
contributions. This mismatch is cured by the
correct calculation.

The calculation proceeds as follows,
\bea
\< \Delta_\mu J^{(5)\mu}_{DW}(x) \>&=& 2 m \< \widetilde q_x \gamma_5 q_x \> \; + \; 2 \< \widetilde Q_{x,L_s/2} \gamma_5 Q_{x,L_s/2+1}\>  \nn
&- &   2  \< \widetilde \Phi_{x,L_s} \gamma_5 \Phi_{x,1}\> +   2 \< \widetilde \Phi_{x,L_s/2} \gamma_5 \Phi_{x,L_s/2+1}\> \,.
\eea
The first term is the usual quark mass contribution. The second gives
two contributions using the identity
\bea
- \< \widetilde Q_{x,L_s/2} \gamma_5 Q_{x,L_s/2+1}\>  &=&  Tr[\gamma_5 A_{L_s/2+1,s}(m)  M_{s,L_s/2}(1)]_{x,x} \\
&=& (1-m) Tr[ \Delta^R_{L_s/2+1} D^{-1}_{ov}(m) \gamma_5 \Delta^L_{L_s/2}]_{x,x} +  Tr[\gamma_5 M_{L_s/2+1,L_s/2}(1)]_{x,x} \nonumber\,,
\eea
where  $Tr[\cdots]$ traces only over color and spin with fixed space time point x.
This contributes both a term corresponding the chiral violation for $m_{res} \ne 0$  and a new term which exactly cancels with the 
mid-term contribution of the Pauli-Villars fields,
\be
- \< \widetilde \Phi_{x,L_s/2} \gamma_5 \Phi_{x,L_s/2+1}\>  =  - Tr[\gamma_5 M_{L_s/2+1,L_s/2}(1)]_{x,x}  \; .
\ee
The change in sign is due to Bose versus Fermi statistics. 
Instead now the anomaly comes for the boundary Pauli-Villars term,
\be
2  \< \widetilde \Phi_{x,L_s} \gamma_5 \Phi_{x,1}\> = \lim_{m\rightarrow 1}
\frac{2}{1-m} Tr[\gamma_5 (D^{ov -1}(m) -1)]_{x,x} = 2 Tr[\gamma_5 (1 - D^{ov}(0) )]_{x,x}
\label{eq:topDensity} \; .
\ee
In the limit of $L_s \rightarrow \infty$, summing over the toroidal volume this gives the lattice
Atya-Singer index for the instanton
number, $(1/2)\sum_x Tr[\gamma_5 D_{ov}(0)]_{x,x} = n_+ - n_- $. 
In the continuum limit, Eq.~\ref{eq:topDensity}  gives the topological charge density $(1/32\pi^2) Tr[F_{\mu,\nu}(x)
\widetilde F_{\mu,\nu}(x)]$ as required~\cite{Niedermayer:1998bi}.
  The Pauli-Villars term is doing its job by
canceling all the heavy cut-off modes in the interior.  
%It makes no
%physical sense to have the anomaly entangled with these modes even
%though one can ``fake'' it  by realizing that also $- 2Tr[\gamma_5
%M_{s,s'}(1)]]$ does give the same expression. (see Appendix~\ref{sec:A})

\subsection{Residual Chiral Violations}
\label{sec:violations}
 
At fixed values of $L_s$ the ``residual mass'' is a common criterion to
measure the magnitude of chiral symmetry violations. This is defined in
Ref.~\cite{Blum:2000kn,Aoki:2002vt} by the correlator 
\be
m_{res}(t)  = \frac{\sum_{\vec x}  \< j_5(\vec x,t,L_s/2)\; j_5(\vec 0,t,L_s) \>_c} 
             { \sum_{\vec x}  \<  \widetilde q_{\vec x,t} \gamma_5 q_{\vec x,t} \;  \widetilde q_0 \gamma_5 q_0  \>_c } 
= \frac{\sum_{\vec x}  \< \widetilde Q_{\vec x,t} \gamma_5 Q_{\vec x,t} \; \widetilde q_0 \gamma_5 q_0  \>_c}
             { \sum_{\vec x}  \<  \widetilde q_{\vec x,t} \gamma_5 q_{\vec x,t} \; \widetilde q_0 \gamma_5 q_0  \>_c } \,,
             \label{eq:MRESCLASSIC}
\ee
in the plateau region with $t$ away from the source and sink. Note
however in our definition of $m_{res}$, we are using our anti-spinors
$\bar q$ in the denominator that removes the unwanted factor
$(1-m)^2$, which of course is irrelevant to the chiral limit ($m =
0$). The fields $\widetilde Q_{\vec x,t} = \widetilde Q_{x,L_s/2} $,
$Q_{\vec x,t} = Q_{x,L_s/2+1}$ are the domain wall fields at the
mid-plane link: $(L_s/2,L_s/2+1)$. The restriction to connected
contributions is a consequence of defining the residual mass via
non-singlet pseudoscalar sources which have no disconnected diagram.

Computing the contractions for the connected diagram and using the
identities in Sec.~\ref{sec:DWcorr} proven in Appendix~\ref{sec:A} for the two
point correlators, $$  \gamma_5 \<   q_0  \widetilde Q_{x}\> \gamma_5 \< Q_{x} \;\widetilde q_0 \>     = D^{\dagger -1}_{ov}(0,y) \rho^{yz}_{L_s}(x) D^{ -1}_{ov}(z,0) \; , $$ 
we have
\be
 m_{res}(t)  = \frac{\sum_{\vec x}Tr[ \<   q_0  \widetilde Q_{x}\> \gamma_5  \< Q_{x} \;\widetilde q_0 \>  \gamma_5]}
             {\sum_{\vec x}Tr[\< \overline q_{x}  q_0  \> \gamma_5 \< q_{x} \; \overline q_0 \> \gamma_5] } = \frac{ \sum_{\vec x}Tr[\rho(x) D^{-1}_{ov} D^{\dagger -1}_{ov} ] }
{ \sum_{\vec x} Tr[D^{-1}_{ov}(x,0) D^{\dagger -1}_{ov}(x,0) ] }
\simeq   \frac{ \<0|  \overline q\gamma_5 \rho(0) q |\pi\>  }{\<0|  \overline q_0\gamma_5 q_0|\pi\>   } \,,
\ee
where,
\be
\rho^{zy}_{L_s}(x) =  \Delta^L_{zx}  \Delta^R_{xy}\,,
\ee
is defined in Eq.~\ref{eq:LRdelta}. The expression on the right in terms of 
the pion to vacuum matrix element holds for large $t$ separating the source and the sink. 

A perhaps less practical but more elegant definition of 
the residual mass is to sum over all time slices,
\be
\overline m_{res}  =  \frac{\sum_{\vec x,t}  \< \overline Q_{\vec x,t} \gamma_5 Q_{\vec x,t} \; \overline q_0 \gamma_5 q_0  \>_c}
             { \sum_{\vec x,t}  \< \overline q_{\vec x,t} \gamma_5 q_{\vec
                 x,t} \; \overline q_0 \gamma_5 q_0  \>_c }   \; ,
\ee
resulting in a form, 
\be
\overline m_{res}=  \frac{Tr[ \Delta_{L_s}(H_5) D_{ov}^{-1} D_{ov}^{\dagger -1}]}{
   Tr\left[ D_{ov}^{-1}  D_{ov}^{\dagger -1}\right]} 
  =         \sum_{\lambda} \; w_\pi(\lambda)\; \Delta_{L_s}(\lambda) \equiv \<\Delta_{L_s}\>_\pi  \; ,
\label{eq:Mres}
\ee
better suited to theoretical analysis.  The sum is is over the spectrum of the eigenvalues of the transfer matrix. Now the trace includes the sum over the spatial index as well.
On the right we introduced the spectral weight for the pion correlator,
$$w_\pi(\lambda) = \frac{ \< \lambda |D_{ov}^{-1}  D_{ov}^{\dagger -1}
|\lambda\>}{\sum_\lambda \< \lambda |D_{ov}^{-1}  D_{ov}^{\dagger -1}
|\lambda\>}  \; . $$
In the limit $m \rightarrow 0$, the improved definition $\overline m_{res}$
is the normalized trace of the violation of the Ginsparg-Wilson relation
written as 
\be
D_{ov}^{\dagger -1}(0) \Delta_{L_s}(H_5)
D_{ov}^{-1} (0)  = ( D_{ov}^{-1}(0) + \gamma_5 D_{ov}^{-1}(0) \gamma_5 -
2)/2  \; .
\ee 
For the polar decomposition the operator $\Delta_{L_s}$ is positive
definite because the approximation always underestimates the sign
function: $|\epsilon_{L_s}(x)| \le 1$.  Consequently, zero residual
mass ($m_{res} = 0$) implies the exact Ginsparg-Wilson relations and unbroken
Ward-Takahashi relations. 

Using this identity (\ref{eq:Mres}), it is  easy to model the residual mass with a reasonable
approximation to the spectral density.  
This model captures well the trends seen in our numerical results in Sec.~\ref{sec:NumTests}. 
%To understand the procedure qualitatively, it is useful to write a closed form
%expression for $m_{res}$ with a reasonable model for the pion
%correlator density function $w_\pi(\lambda)$ as a function of the
%parameters.  We illustrate this briefly in our analysis of the
%M\"obius performance study in Sec.~\ref{sec:NumTests}
%For example we have been able to model our residual mass for our
%scaled Shamir parameterizations $H_5 = \alpha \gamma_5 D^{Shamir}(M_5)$.
For the spectral density, $w(\lambda)$, we note that
it is plausible that it will have negligible dependences on
$L_s$ and $\alpha$ parameterized in terms of unscaled eigenvalues of
$H_5 = \gamma_5 D^{Shamir}(M_5)$ so that 
\be
m_{res} \simeq \sum_\lambda w_\pi(\lambda) \Delta_{L_s}(\alpha \lambda) \; .
\ee
A simple model for $w_\pi(\lambda)$ is to approximate
a few smallest eigenvalues by a finite density $\rho(0)$ due to ``small topological defects'' and at larger eigenvalues by the  free
kernel.  Indeed this ansatz is able to give a good fit to our empirical study of the 
parameter dependence of $m_{res}$ in Sec.~\ref{sec:NumTests}.  

At large $L_s$ the defect dominate the contribution to $m_{res}$.  Using the expression $\epsilon_{L_s}(\alpha \lambda) = \tanh(x)$ with
$x = L_s \log(1 + \alpha \lambda) -  L_s \log(1 - \alpha \lambda))$ we get the expression for the
error 
\be
\Delta_{L_s}(\alpha \lambda) = 1/(4\cosh^2(x))  \rightarrow e^{- \textstyle  L_s |\log(1+\alpha \lambda) - \log(1-\alpha \lambda)|} \; ,
\ee
for $ O(L_s^{-1}) < | \alpha \lambda| < O(L_s)$. Outside this window the error is $O(1)$.  
For large $L_s$ the error is dominated by
the small eigenvalues in the interval $|\lambda| < 1/(\alpha L_s)$. For the standard 
Shamir form ($\alpha = 1$) this causes the residual to fall like  $m_{res} \sim \rho(0)/L_s$ asymptotically but
for the rescaled M\"obius with $\alpha  \sim L_s/\lambda_{max}$ we estimate
the residual to fall like  $m_{res} \sim \rho(0)/L^2_s$ asymptotically~\footnote{Strictly speaking one should be careful about the order of limits. 
The eigenvalue distribution
is only properly given by a density function,  $\rho(\lambda)$, in the limit of infinite lattice volume  ($L^4$), so here at  finite volume $\rho(0)$ should be replaced by a  measure of the mean number of eigenvalues in the
interval near $\lambda = 0$ where the exponential approximation to $\epsilon(\lambda)$ fails.
At fixed volume, $L^4$, exponential convergence to $m_{res} = 0$ is expected to  resume for $L_s  \gg  L $ but this is of little practical  importance.}. This improvement is,
we believe, the basic explanation for the superior chirality of the M\"obius algorithm.

As we mentioned above, the polar approximation to the sign function results in a positive  residual mass.
However, other polynomial approximations,  such as Zolotarev, of the sign function may oscillate around $\epsilon(x)$ so
positivity is lost. In fact one can even ``tune'' $m_{res}$ to zero
but this does {\bf not} imply that chiral symmetry is exact.  Instead
it implies that the lowest order dimension 3 operator in to the
chiral Lagrangian is given by the quark mass. The real issue is higher
order terms, in particular the dimension 5 chiral symmetry breaking
operators.

\subsection{Baryon current and Chemical Potential}

The chemical potential couples to the singlet charge of the vector
current, i.e.\ Baryon number.  In this section we point out the relation of the vector current we already defined to the non-zero chemical
potential formulation of the overlap and domain wall Fermions. It is useful to note that the same strategy can be used to
define currents for kernels that violate $\gamma_5$ Hermiticity
leading to complex determinants such as the Dirac operator with a
chemical potential.  Block and Wettig~\cite{Bloch:2006cd} have given
the overlap operator for non-zero chemical potential. Their
construction is tricky because of the need to define the ``sign
function'' for a non-Hermitian kernel $''\epsilon( A)''$, where
$A=\gamma_5 D^{Wilson}(M_5, \mu)$ is given by making the standard
substitution for non-zero chemical potential,
\be
(1 \pm  \gamma_4) U_{\pm 4}(x) \rightarrow  (1 \pm \gamma_4) e^{\pm \mu} U_{\pm 4}(x)
\ee
into the Wilson operators for all time like links. Their rule
is to define   the ``sign function'' by
\be
``\epsilon(A)'' = \;  S\mbox{sign} (Re \; \Lambda) S^{-1}\,,
\ee
where $S$ is the similarity transformation that
diagonalizes the kernel $A = S \Lambda S^{-1}$. 

It is straight forward to rederive their prescription and generalize
it for any kernel at finite $L_s$ by simply inserting this Wilson
operator (for non-zero $\mu$) into our generalized domain wall formalism.
In the domain wall form we get,
\be
\epsilon_{L_s}(A) =\frac{T^{-{L_s}} -1}{T^{-{L_s}} + 1} = \frac{(1+A)^{L_s} - (1-A)^{L_s}}
{(1+A)^{L_s} + (1-A)^{L_s}}    = S \frac{(1+\Lambda)^{L_s} - (1-\Lambda)^{L_s}}
{(1+\Lambda)^{L_s} + (1-\Lambda)^{L_s}} S^{-1}   \,,
\ee
where we diagonalize $A = S \Lambda S^{-1}$ with eigenvalues $\lambda
= x +i y$.  As shown in Appendix~\ref{sec:A}, $A = (T^{-1} +1)^{-1} (T^{-1} -1) 
= \gamma_5 (D_+ + D_-)/[\gamma_5(D_+ - D_-)
\gamma_5]$ or $A = \gamma_5 D_W(\mu)$ for Bori\c{c}i. But using $|(1 \pm
\lambda)^{L_s}| = |1 \pm \lambda|^{L_s}$ or $|(1 \pm x \pm i y)^{L_s}|
= [(1 \pm x)^2 + y^2]^{{L_s}/2}$ we see that the exponentially
dominant of the two terms for each eigenvalue is the one where the
sign of $\pm x$ is positive.  Hence we get
\be
\epsilon_{L_s}(A) \rightarrow S \mbox{sign}(Re \Lambda) S^{-1}\,,
\ee 
in {\bf exact} agreement with Block and Wettig at $L_s  = \infty$.

\newpage
%%%%%%%%%%%%%%%%%%%%%%%%%%%  Section %%%%%%%%%%%%%%%%%%%%%%%%%%%%%%%%

\section{Conclusions}
\label{sec:conclusions}

In this paper we have reviewed the M\"obius class of chiral domain
wall Fermion operators~\cite{Brower:2004xi,Brower:2005qw}. Since they
are just now coming into wider use in production codes, a general
presentation of the formalism is perhaps warranted. We have sought to
emphasize several features critical to their performance and chiral
properties. On the performance side, we note that the M\"obius kernel
operator requires no additional applications of the 4-d Wilson Dirac
kernel per domain wall iteration. However we are required to replace
the conventional 5-d red-black preconditioning by a 4-d checkerboard
with constant color in the 5-th axis in order to avoid a 4-d Wilson
operator inverse as a new inner loop. In fact such 4-d even/odd
decomposition is needed for the earlier Bori\c{c}i domain
wall formulation~\cite{Borici:1999da,Borici:1999zw}, as well as any
operator that goes beyond nearest neighbor terms in the 5-th direction. A
highly optimized parallel M\"obius Domain Wall Fermion (MDWF) inverter
for clusters and the Blue Gene architecture has been freely available
for several years~\cite{Pochinsky:2008zz} as well as Hybrid Monte
Carlo evolution code in Chroma~\cite{Edwards:2004sx}. Very soon the
M\"obius inverters will be available in the QUDA ~\cite{Clark:2009wm}
(QCD in CUDA) library for NVIDIA multi-GPUs platforms as well as a
full Hybrid Monte Carlo evolution code in the Columbia Physics System
~\cite{CPS:2012} (CPS) for M\"obius fermions optimized for the
BlueGene/Q.

In addition, in this paper we worked out the form of the conserved
and partially conserved axial vector currents for M\"obius domain wall fermions, as well as their
Ward-Takahashi identities. The goal was to have a general approach that maps the domain
wall expressions at finite $L_s$ into their equivalent form for the effective
4-d overlap action. A byproduct of our formalism is a simple derivation of the
overlap operator at finite chemical potential, a result first obtained by Block and Wettig~\cite{Bloch:2006cd}.  Also we  show how the general expression for the
residual mass, in the case of non-vanishing zero mode density,   implies quadratic convergence,  $m_{res} = O(1/L^2_s)$ , for the appropriately scaled M\"obius fermions for
large $L_s$ in contrast to the slower linear convergence,  $m_{res} = O(1/L_s)$, for Shamir. This largely
explains the reason for the improved chiral behavior of the M\"obius  rescaling algorithm.
Indeed,  for the test ensemble we used, our numerical tests support this picture and even suggest 
 that M\"obius at $L_s = 32$ should correspond roughly to running Shamir with $L_s = {\cal O}(10^3)$. 
Of course the latter  is neither practical or even easily amenable to direct numerical 
verification.

\begin{figure}[h]
\begin{center}
\includegraphics[width=0.6\textwidth]{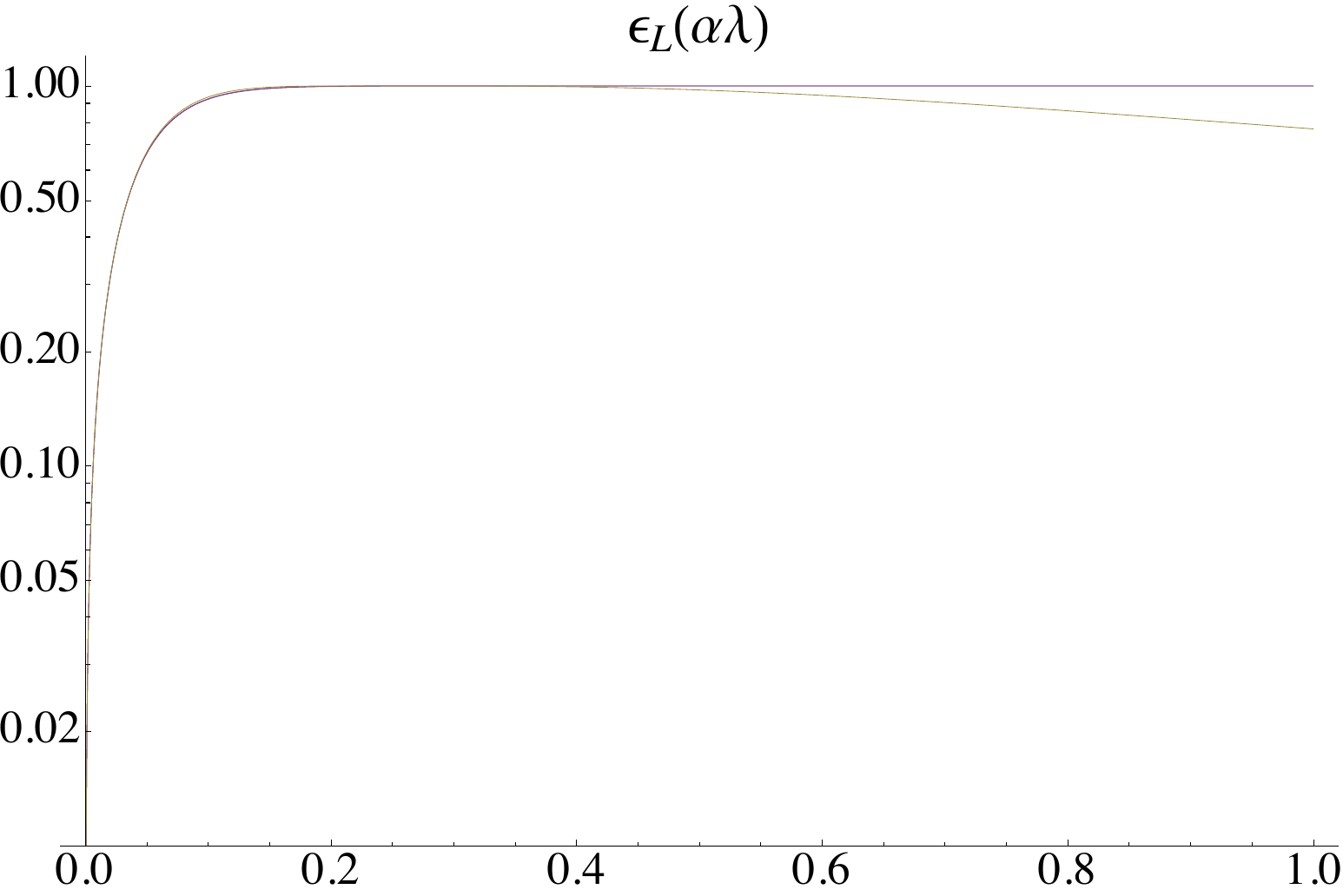}
\caption{  Comparison the approximate of the sign function $\epsilon_{L_s}(\alpha \lambda)$ plotted
against $\lambda$  for
Shamir ($\alpha = 1$) at $L_s = 16$   with rescaled M\"obius ($\alpha = 2$) at $ L_s = 8$   (indistinguishable) and
the lower curve scaled M\"obius ($\alpha = 4$) at $L_s = 4$ . The support for the kernel is bounded by eigenvalues $|\lambda| < (8-M_5)/(10- M_5)  \simeq 0.75 $ so the visible degradation at the top is limited and should  not be significant  for the chiral physics at small eigenvalue.}
\label{Fig:scaling16}
\end{center}
\end{figure}

\begin{figure}[h]
\begin{center}
\includegraphics[width=0.8\textwidth]{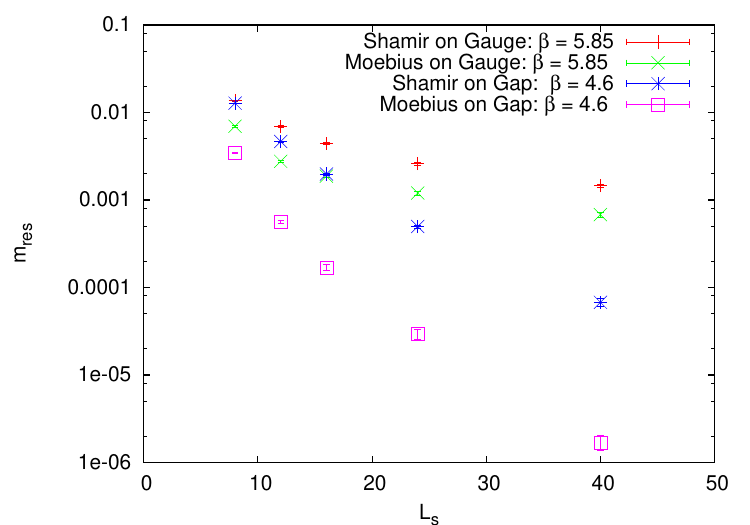}
\caption{ The M\"obius algorithm ($\alpha = 2$) on pure gauge and
Gap lattices vs Shamir ($\alpha = 1$).}
\label{Fig:moebius_gap}
\end{center}
\end{figure}

Perhaps of more interest is to  consider using the M\"obius algorithm to reduce $L_s$ substantially.
For  new HMC runs, one can  be  even more aggressive. For example as 
illustrated in Fig.~\ref{Fig:scaling16}, the
typical simulation with Shamir at $L_s = 16$ might be run with M\"obius at $L_s$ as low as $L_s = 4$ with a tolerable
 compromise on chirality. Perhaps not all simulations can be this aggressive but
there  may be some  that can. For example exploratory investigations for Beyond the Standard Model (BSM) physics might
try this at least in early broad surveys of the gauge theory landscape.

We  are also aware that as the M\"obius algorithm comes into more
common use, there are potentially opportunities to combine it with
other algorithmic methods  such as Hasenbusch~\cite{Hasenbusch:2001ne}
mass precondition or multigrid
preconditioning~\cite{Babich:2010qb, Babich:2009pc}. For
the latter it is interesting  to note that the  first level of domain wall multigrid maps the 5-d operator into an
effective 4-d coarse operator~\cite{Cohen:2012sh} suggesting many
avenues of investigations that could allow domain wall codes to
compete favorably with 4-d Wilson codes. One example of a hybrid
algorithm that has received some preliminary investigation is combing
the M\"obius algorithm with the idea of Gap
Fermions~\cite{Vranas:2006zk} to further reduce the value of $m_{res}$
at fixed $L_s $ as noted in Fig.~\ref{Fig:moebius_gap} reproduced from
Ref.~\cite{Brower:2009sb}. The figure shows that the smoothing
consequences of introducing a Gapped action can further reduce the
residual mass  by as much as an order of magnitude for $L_s
= 16$ for example. 

The interaction between different algorithmic methods opens up
a large range of possible improvements worth of serious study. The RBC
collaboration has been pursuing similar strategies~\cite{Kelly:2012uy}, to try to balance
the apparently conflicting desire for better chirality while still
facilitating sufficient thermalization of topological charge sectors.
We have no magic solution to this but of course the interaction of
M\"obius fermions with  other algorithmic approaches begs further
explorations in this endeavor.

\paragraph{Acknowledgment:}
We wish to acknowledge many useful discussions and  interactions  with Andrew Pochinsky, Claudio Rebbi, David Schaich and Pavlos Vranas.  This research was supported in part at William and Mary by DOE grant DE-FG02-07ER41527 and  by DE-AC05-06OR23177 (JSA),  and in part at Boston University by U.S.\ DOE grants DE-FG02-91ER40676 and DE-FC02-06ER41440; NSF grants DGE-0221680, PHY-0427646, and PHY-0835713.

\newpage
\bibliographystyle{unsrt}
\bibliography{moebius}

\pagebreak

%%%%%%%%%%%%%%%%%%%%%%%%%%%%%% APPENDIX %%%%%%%%%%%%%%%%%%%%%%%%%%

\appendix
\section{M\"obius generalization of Domain Wall Operator}
\label{sec:A}

This appendix collects together the basic identities that relate our
generalized domain wall operator at finite $L_s$ to the effective operators in the 4d
overlap world.   The class of domain wall operator (including our
M\"obius and Zolotarev cases)  has next to nearest neighbor interaction in the fifth
axis,
\bea
D_{DW}(m)_{s,s'}  &=&  
  D^{(s)}_- \, P_+ \,\delta_{s,s'+1} + D^{(s)}_+ \, \delta_{s,s'} 
+   D^{(s)}_- \, P_-  \, \delta_{s,s'-1}    \nn
&-& m \, D^{(1)}_- P_+ \, \delta_{s,1}\delta_{s',L_s}  - m \,  D^{(L_s)}_- P_-
\, \delta_{s,L_s}\delta_{s',1} \,,
\eea
 with $s,s' = 1, 2, \cdots L_s$
mod $L_s$ cyclic moduls $L_s$ and $P_\pm = \Half(1 \pm \gamma_5)$ and $D^{(s)}_+ = b_5(s) D^{Wilson}(M_5)
+1$, $D^{(s)}_- = c_5(s) D^{Wilson}(M_5) -1$. The Wilson operator is
\be
D^{Wilson}_{xy}(M_5) =  (4+M_5) \delta_{x,y} - 
 \frac{1}{2} \Bigl[  (1 - \gamma_\mu) U_\mu(x) \delta_{x+\mu,y} 
+  (1 + \gamma_\mu) U_\mu^\dagger(y) \delta_{x,y+\mu} \Bigr] \; . 
\label{eq:Waction} 
\ee
In matrix notation the domain wall operator is
\be \label{eq:DWmatrix}
D_{DW}(m) =
\begin{bmatrix}
D^{(1)}_+ & \quad D^{(1)}_- P_- & 0 &\cdots &  -mD^{(1)}_- P_+  \\
\quad D^{(2)}_- P_+ & D^{(2)}_+ & \quad D^{(2)}_- P_- &\cdots & 0  \\
0 & \quad D^{(3)}_- P_+ & D^{(3)}_+ & \cdots&   0  \\
\vdots & \vdots & \vdots &  \ddots & \vdots \cr
-m D^{(L_s)}_- P_- & 0 & 0  &  \cdots & D^{(L_s)}_+  \\
\end{bmatrix}
\; .
\ee
We will also on occasion discuss the equally valid ``left'' form
\be
\widehat D_{DW}(m)  = D^{-1}_-  D_{DW}(m) D_- \,,
\ee
with $D_- = Diag[D^{(1)}_- , D^{(2)}_- , D^{(2)}_- ,\cdots , D^{(L_s)}_-]$,
which rotates the chiral projectors to the left,
\be
\widehat  D_{DW}(m) =
\begin{bmatrix}
D^{(1)}_+ & \quad P_-  D^{(2)}_- & 0 &\cdots &  -m P_+ D^{(L_s)}_-  \\
\quad P_+ D^{(1)}_-  & D^{(2)}_+ & \quad  P_- D^{(3)}_- &\cdots & 0  \\
0 & \quad P_+  D^{(2)}_- & D^{(3)}_+ & \cdots&   0  \\
\cdots & \cdots & \cdots & \cdots &  \cdots  \\
-m P_- D^{(1)}_- & 0 & 0  &  \cdots & D^{(L_s)}_+  \\
\end{bmatrix}\,,
\label{eq:leftDW}
\ee
because of its possible advantage for efficient code. 

\subsection{LDU decomposition}

To relate the 5d domain wall matrix to the 4 dimensional overlap form, one
merely performs a standard {\bf LDU decomposition} of Eq.~\ref{eq:DWmatrix},
although the notation in the domain wall literature (which we adhere to here)
is unconventional and obscures this a bit. Consider the decomposition,
\be
D_{DW} {\cal P} = ``U D L" \; .
\ee
The interchange of U and L relative to conventions of mathematics texts is
inconsequential since reflecting the 5-th axis with transformation, ${\cal R}$  defined in Eq.~\ref{eq:Rsym} converts any $U$ matrix to an $L$ matrix and
vice versa.

The first step is to multiply by the permutation (or pivot) matrix 
that performs a left shift $C$ for positive chirality 
components:
\be \label{eq:Pmatrix}
{\cal P} = P_- I + P_+ C=  
\left[\begin{array}{ccccc}
P_- & P_+  & 0 & \cdots & 0  \\
0 & P_-  & P_+ & \cdots & 0  \\
0 & 0  & P_- & \cdots   & 0  \\
\cdots & \cdots & \cdots & \cdots &  \cdots  \\
P_+ & 0  & 0 \cdots & & P_- \\
\end{array} \right]\,.
\ee
With $Q^{(s)}_- = \gamma_5[D^{(s)}_- P_+ + D^{(s)}_+ P_-]$ and $Q^{(s)}_+ =
\gamma_5[D^{(s)}_+ P_+ + D^{(s)}_- P_-]$, 
this gives an upper diagonal form:
\be \label{eq:Dpivot}
 D_{DW}(m) \, {\cal P} =   \gamma_5
 \begin{bmatrix}
 Q^{(1)}_- c_-  & Q^{(1)}_+ & 0 & \cdots  & 0  \\
 0 &  Q^{(2)}_- & Q^{(2)}_+ & \cdots & 0  \\
 0 & 0  &  Q^{(3)}_- & \cdots & 0  \\
\vdots & \vdots & \vdots & \ddots & \vdots  \\
 Q^{(L_s)}_+ c_+  & 0  & 0 & \cdots &  Q^{(L_s)}_-  \\
\end{bmatrix}\,,
\ee
or,
\be
 D_{DW}(m) \, {\cal P} =  \gamma_5 Q_-
\begin{bmatrix}
 c_- & -T^{-1}_1 & 0 & \cdots  & 0  \\
 0 &  1 & -T^{-1}_2 & \cdots & 0  \\
 0 & 0  &  1 & \cdots & 0  \\
\vdots & \vdots & \vdots & \ddots & \vdots  \\
-T^{-1}_{L_s} c_+ & 0  & 0 & \cdots &  1  \\
\end{bmatrix}\,,
\ee
after factoring out the the diagonal matrix, $Q_- = Diag[Q_-^{(1)}, Q_-^{(2)}, \cdots,Q_-^{(Ls)}]$.
Here we define the mass dependent constants, $c_\pm= P_\pm - m P_\mp = \half (1-m) \pm \half (1+m) \gamma_5$, and the 4d local ``transfer matrix'',  $T_{[s+1,s]} \equiv T_s$, from s to s+1 
on the $[s+1,s]$ or in fact its inverse,
\be
T_{[s,s+1]} \equiv T_{s}^{-1}  = - (Q_-^{(s)})^{-1} Q_+^{(s)}  \; .
\label{eq:Tmatrix}
\ee
This operator is Hermitian and relates to a 5d s dependent Hamiltonian operator by the identity,
\be
H_s = \frac{T^{-1}_{s} - 1}{ T^{-1}_{s} + 1} =  \frac{1}{Q^{s}_+ - Q^{s}_- }  
(Q^{s}_+ + Q^{s}_-) = \gamma_5 \frac{D^{(s)}_+ + D^{(s)}_-}{D^{(s)}_+ - D^{(s)}_-}\,.
\ee
Thus we have an s dependent l version of the M\"obius operators given in 
Eq.~\ref{eq:Moebius},
\be
H_s = \gamma_5 \frac{(b_5(s) + c_5(s))D^{Wilson}(M_5)}{2 +  (b_5(s) - c_5(s)) D^{Wilson}(M_5)} \; .
\ee
The remaining two steps are Gaussian elimination with ``U''
 and back substitution with ``L'',
to obtain ~\footnote{Again we point out that the notation is a little
  unconventional relative to the mathematics literature, since we
  should really identify the ``U'' matrix as $\gamma_5 \, Q_- \, U\,
  Q^{-1}_- \gamma_5$, absorbing $\gamma_5 \, Q_- \,$ into the diagonal
  matrix.}
\begin{equation}
  D_{DW}(m) \, {\cal P}  =  \gamma_5  \, Q_- \, U  D_5(m) \, L(m)\,,
\end{equation}
in terms of
\be
U= \begin{bmatrix}
1 & - T_1^{-1}  & 0 &  \cdots & 0 \\
0 & 1        & - T_2^{-1}          & \cdots & 0   \\
0 & 0        & 1                & \cdots &   0   \\
\vdots & \vdots & \vdots & \ddots & \vdots  \\
0 & 0        & 0                & \cdots & 1 \\
\end{bmatrix}
\qquad
L(m) = 
\begin{bmatrix}
- 1 & 0  & 0 &\cdots & 0  \\
-T_{[2,1]} c_+ & 1  &  0 & \cdots & 0  \\
-T_{[3,1]}  c_+ & 0  & 1 & \cdots & 0  \\
\vdots & \vdots & \vdots & \ddots & \vdots  \\
-T_{[Ls,1]}  c_+ & 0  & 0 & \cdots & 1 \\
\end{bmatrix}\,,
\ee
where  $T_{[s,1]} = T^{-1}_s T^{-1}_{s+1} \cdots T_{L_s}^{-1}$, assuming
periodic index notation  $1 = \mbox{\bf mod} L_s +1 $. For future reference this notation
is generalized to s-ordered products ( $1  \le s < s' \le L_s $),
\be
T_{[s,s'+1]} = T^{-1}_{s} T^{-1}_{s+1} \cdots T^{-1}_{s'}  \quad , \quad 
 T_{[s'+1,s]} = T_{s'}  \cdots T_{s+1} T_{s} \,,
\ee
with  special cases:
$T_{[s,s]} = 1 = $, $T_{[s+1,s]} = T_s  $,$ T_{[s,s+1]}= T^{-1}_s $  and 
\be
{\mathbb T}^{L_s} \equiv   T_{L_s} \cdots T_3 T_2 T_1  \quad \mbox{and} \quad
{\mathbb T}^{-L_s} \equiv  T_1^{-1}T_2^{-1}T_3^{-1} \cdots T_{L_s}^{-1} \; . 
\ee
We also introduced the matrix: $D_5(m) = Diag[D_4(m), 1, \cdots,1]$, where $D_4(m)  \equiv {\mathbb T}^{-L_s}c_+  - c_-$, 
\be
D_4(m)  =  \frac{1+m}{2} ({\mathbb T}^{-L_s}+1)\gamma_5  
+ \frac{1-m}{2} ({\mathbb T}^{-L_s}-1 ) = 
[({\mathbb T}^{-L_s}+1)\gamma_5]\times[ \frac{1+m}{2} + \frac{1-m}{2} \gamma_5 \frac{{\mathbb T}^{-L_s}-1}{{\mathbb T}^{-L_s}+1}] \,,
\ee
The inverses are $L^{-1}(m) = L(m)$  and
\be
U^{-1}= \begin{bmatrix}
1 & T_{[1,2]}  & T_{[1,3]} &  \cdots &  T_{[1,L_s]} \\
0 & 1        & T_{[2,3]}         & \cdots &  T_{[2,L_s]}   \\
0 & 0        & 1                & \cdots &   T_{[3,L_s]}   \\
\vdots & \vdots & \vdots & \ddots & \vdots  \\
0 & 0        & 0                & \cdots & 1 \\
\end{bmatrix}\,.
\ee
The only non-trivial diagonal element in $D_5(m)$ can be factored 
\be
 D_{ov}(m) \equiv D^{-1}_4(1)D_4(m) = \frac{1+m}{2} + \frac{1-m}{2} \gamma_5 \epsilon_{L_s}[{\mathbb H}]\,,
\ee
where
\be
\epsilon_{L_s}[{\mathbb H}] \equiv  \frac{{\mathbb T}^{-L_s} -1}{{\mathbb T}^{-L_s} + 1}  \,,
\ee
and ${\mathbb H} = (1 -  {\mathbb T})/(1 + {\mathbb T})$. Gamma 5 Hermiticity requires
that ${\mathbb H}$ and therefore that $\mathbb T$   is Hermitian and therefore,
$T_1 T_2 T_3 \cdots T_{L_s}  = T_{L_s} \cdots T_{3} T_{2} T_{1}$.

It is now straight
forward to compute the matrix  in Eq.~\ref{eq:DWFmatrix}
\be
 {\cal P}^{\dagger} \frac{1}{D_{DW}(1)}D_{DW}(m) {\cal P} = L(1)
 \; Diag[D_{ov}(m), 1, \cdots, 1] \; L(m)\,,
\ee
and its inverse in Eq.~\ref{eq:inverseDWFmatrix},
\be
A =  {\cal P}^{\dagger} \frac{1}{D_{DW}(m)}D_{DW}(1) {\cal P} = L(m)
 \; Diag[D^{-1}_{ov}(m), 1, \cdots, 1] \; L(1)\,,
\ee
where 
\begin{equation}
  A^{DW}_{ss'} = 
 \left[ \begin{array}{rrrrrrr}
D^{-1}_{ov}(m) & 0 & 0 & \cdots & \cdots & \cdots& 0\\
(1-m)\Delta^R_{2} D^{-1}_{ov}(m)  & 1 &0 &0&\cdots &\cdots& 0\\ 
(1-m)\Delta^R_{3} D^{-1}_{ov}(m)  & 0 &1 & 0 & \cdots &\cdots&0\\
(1-m)\Delta^R_{4}  D^{-1}_{ov}(m)  & 0 & 0 & 1 & \cdots & \cdots & 0 \\
\vdots & \vdots &  \ddots & \ddots &\ddots &\ddots &\vdots \\
(1-m) \Delta^R_{L_s}  D^{-1}_{ov}(m)  & 0 & \cdots &\cdots&\cdots&0& 1
\end{array}\right]  \; .
\end{equation}
Finally we have the crucial identity that relates the 5d and 4d
determinants,
\be
Det[D^{DW -1}(1) D_{DW}(m)] = Det[D_{ov}(m)] \; ,
\ee
because the $\gamma_5 Q_- U$ factor cancels in the product and $- Det[L(1)] =
- Det[L(m)] = 1$.  This completes the proof that the overlap measure is
equivalent to the domain wall Fermion measure with Pauli-Villars pseudo-Fermion 
field to give the factor $Det[D^{DW -1}(1)]$.

\subsection{Domain Wall Correlators}
\label{sec:DWcorr}

The general bulk to bulk propagator, Eq.~\ref{eq:inverseDWFmatrix},
\be
\< \Psi_s \overline \Psi_s \> \equiv D^{-1}_{DW}(m)_{s,s'}\,,
\ee
is the domain wall inverse itself.  For the Ward-Takahashi identities we need also to have boundary to bulk
propagators to interior points in the domain wall.
One fundamental set is 
\bea
\<q \overline \Psi_{s'} \> &=& [{\cal P}^\dagger D^{-1}_{DW}(m)]_{1s'} 
= P_- D^{-1 DW}_{1s'}(m) + P_+ D^{-1 DW}_{L_ss'} (m) \\
\<\Psi_s \overline q \> &=& [ D^{-1}_{DW}(m)D_{DW}(1){\cal P}]_{s1}\,.
\eea
In addition we need bulk to bulk propagators between  quark spinors 
\be
Q_{s} = [{\cal P}^\dagger \Psi]_{s} = P_- \Psi_{s} + P_+\Psi_{s-1}
\ee
and two varieties of anti-quark spinors
\bea
\overline  Q_s &=& [\overline  \Psi  \, D_{DW}(1)  {\cal P}]_{s} 
\nonumber \\
\widetilde Q_s &=& [\overline  \Psi  \, (- D_-) {\cal P^\dagger}]_{s} 
=  \overline \Psi_{s} (-  D^{(s)}_-) P_- + \overline \Psi_{s+1} (- D^{(s + 1)}_-)  P_+  \,.
\eea
Thus we introduce the two correlators. The first, 
\be
A^{DW}_{ss'}(m) = \<Q_s \overline  Q_{s'}\> = [{\cal P}^\dagger \frac{1}{D^{DW}(m)}D^{DW}(1){\cal P}]_{s,s'}\,,
%\label{eq:Amatrix}
\ee
is the fundamental formula  derived above.  The second ``mass'' term,
\be
M_{s,s'}(m) = \<Q_s \widetilde Q_{s'}\> = [{\cal
    P}^\dagger \frac{1}{D^{DW}(m)}( -D_-){\cal P^\dagger}]_{s,s'}\,,
%\label{eq:Matrix}
\ee
is needed for the axial.  There is also  a very convenient factorization  formula,
$M(m) = A^{DW}(m) M(1)$, for this.

For the {\bf flavor} chiral ward identity two correlators are
needed
\bea
A^{DW}_{s,1} = \<Q_s \overline q\> &=& (1-m) X_{s-1} D^{-1}_{ov}(m) =   
(1-m) T^{-1}_{s} \cdots T^{-1}_{L_s}   \frac{1}{1 +
  {\mathbb T}^{-L_s} }  D^{-1}_{ov}(m) \nn
&\equiv& (1-m) \Delta^R_s  D^{-1}_{ov}(m) \,.
\label{eq:Qqcorr}
\eea
The second one is
\be
M_{1,s} = \<q \widetilde Q_s \> =  \gamma_5 D^{\dagger -1}_{ov}(m) \frac{1}{1 +  {\mathbb T}^{-L_s} } 
 [T^{-1}_{1} \cdots T^{-1}_{s}]\gamma_5 \equiv  D^{-1}_{ov}(m) \gamma_5
 \Delta^L_s \gamma_5\,.
\label{eq:qQcorr}
\ee
This follows from the reflection property $T^{-1}_s = T^{-1}_{L_s +1 -s}$, the identity $\overline q = (1-m) \widetilde q +
[\overline \Psi D_{DW}(m) {\cal P}]_1$, which
implies
\be
M_{s,1} = \< Q_s \widetilde q \>  =  \Delta^R_s  D^{-1}_{ov}(m)  + \delta_{s,1} \,,
\ee
and the fact that $ {\cal R} D^{-1}_- D^{DW}(m)$ is $\gamma_5$
Hermitian, so that $ \< q \widetilde Q_s\> = \gamma_5 {\cal
  R}_{ss'}\<Q_{s'} \widetilde q \>^\dagger \gamma_5 $.

Finally for the chiral anomaly in the {\bf singlet} current, we need the correlator 
\bea
M_{ss'}(m) &=& \< Q_s \widetilde Q_{s'} \> =  A^{DW}_{ss''}(m) M_{s'', s'}(1) = (1-m) \Delta^R_s D^{-1}_{ov}(m) 
M_{1,s'}(1)  + M_{s,s'}(1)  \nn
&=& (1-m) \Delta^R_s D^{-1}_{ov}(m) \gamma_5 \Delta^L_{s'} \gamma_5 +  M_{s,s'}(1)\,,
\eea
for the anomaly with $s = L_s/2+1, s' = L_s/2$.

\end{document}